\documentclass[9pt,twocolumn,twoside]{pnas-new}

\templatetype{pnasresearcharticle} 
\begin{document}

\title{Quantum Benchmarking of High-Fidelity Noise-Biased Operations on a Detuned-Kerr-Cat Qubit}

\author[a, 1, 2]{Bingcheng Qing}

\author[a, 1]{Ahmed Hajr}

\author[a,b,2]{Ke Wang}

\author[a]{Gerwin Koolstra}

\author[a,b,2]{Long B. Nguyen}

\author[a]{Jordan Hines}

\author[c]{Irwin Huang}

\author[d]{Bibek Bhandari}

\author[a]{Larry Chen}

\author[a]{Ziqi Kang}

\author[a,b]{Christian Jünger}

\author[a]{Noah Goss}

\author[a]{Nikitha Jain}

\author[a]{Hyunseong Kim}

\author[a,b]{Kan-Heng Lee}

\author[a,b]{Akel Hashim}

\author[e]{Nicholas E. Frattini}

\author[a,b]{Zahra Pedramrazi}

\author[d,f]{Justin Dressel}

\author[c,d,f,g]{Andrew N. Jordan}

\author[a,b]{David I. Santiago}

\author[a,b]{Irfan Siddiqi}

\affil[a]{Quantum Nanoelectronics Laboratory, University of California, Berkeley, CA, 94720, USA}

\affil[b]{Applied Mathematics and Computational Research Division, Lawrence Berkeley National Laboratory, Berkeley, CA, 94720, USA}

\affil[c]{Department of Physics and Astronomy, University of Rochester, Rochester, NY, 14627, USA}

\affil[d]{Institute for Quantum Studies, Chapman University, Orange, CA, 92866, USA}

\affil[e]{Department of Applied Physics and Physics, Yale University, New Haven, CT, 06520, USA}

\affil[f]{Schmid College of Science and Technology, Chapman University, Orange, CA, 92866, USA}

\affil[g]{The Kennedy Chair in Physics, Chapman University, Orange, CA, 92866, USA}

\leadauthor{Qing}


\significancestatement{The path to practical quantum computing is hindered by the presence of noise, which disrupts fragile quantum information. Recent advances suggest that tailoring quantum hardware to favor certain types of noise can lead to more efficient error correction strategies. In this study, we develop and rigorously benchmark a scalable control framework for the detuned Kerr-cat qubit, a leading noise-biased platform. Our results set a new performance benchmark, demonstrating unprecedented levels of noise bias and gate fidelity. Importantly, we reveal that conventional estimates often overstate the noise in these systems, underscoring the need for more precise diagnostic tools. This work brings us closer to fault-tolerant quantum computing by advancing both the hardware and the methodology needed to unlock its full potential.}

\authorcontributions{}
\equalauthors{\textsuperscript{1}B.Q. contributed equally to this work with Ah.H.}
\correspondingauthor{\textsuperscript{2}To whom correspondence should be addressed. Electronic addresses: bc.qing@berkeley.edu, kywang@berkeley.edu, longbnguyen@berkeley.edu}



\begin{abstract}
Ubiquitous noises in quantum systems remain a key obstacle to building quantum computers, necessitating the use of quantum error correction codes. Recently, error-correcting codes tailored for noise-biased systems have been shown to offer high fault-tolerance thresholds and reduced hardware overhead, positioning noise-biased qubits as promising candidates for building universal quantum computers. However, quantum operations on these platforms remain challenging, and their noise structures have not yet been rigorously benchmarked to the same extent as those of conventional quantum hardware. In this work, we develop a comprehensive quantum control toolbox for a scalable noise-biased qubit, detuned Kerr-cat qubit, including initialization, universal single-qubit gates and quantum non-demolition readout. We systematically characterize the noise structure of these operations using gate set tomography and dihedral randomized benchmarking, achieving high local gate fidelities, with $\mathcal{F}[Z(\pi/2)]=99.2\%$ and $\mathcal{F}[X(\pi/2)]=92.5\%$. Notably, the noise bias of the detuned Kerr-cat qubit approaches 250, which outperforms its resonant-Kerr-cat qubit counterparts as reported previously, representing a new state-of-the-art performance benchmark for noise-biased qubits. Moreover, our results reveal a critical overestimation of operational noise bias inferred from bit-flip and phase-flip times alone, highlighting the necessity of a precise and direct benchmarking for noise-biased qubit operations. Our work thus establishes a framework for systematically characterizing and validating the performance of quantum operations in structured-noise architectures, which lays the groundwork for implementing efficient quantum error correction in next-generation architectures.
\end{abstract}

\dates{This manuscript was compiled on \today}
\doi{\url{www.pnas.org/cgi/doi/10.1073/pnas.XXXXXXXXXX}}

\maketitle
\thispagestyle{firststyle}
\ifthenelse{\boolean{shortarticle}}{\ifthenelse{\boolean{singlecolumn}}{\abscontentformatted}{\abscontent}}{}

\firstpage[4]{2}

\dropcap{R}apid advancements in building quantum systems with scalable, well-characterized qubits have opened up the potential to solve problems intractable for classical computers~\cite{shor1999polynomial, arute2019quantum,zhong2020quantum}. However, quantum systems remain fragile, subject to various errors such as decoherence~\cite{martinis2005decoherence}, dephasing~\cite{bluvstein2024logical, reilly2008suppressing}, stochastic and coherent errors~\cite{hashim2020randomized}. As a result, quantum error correction (QEC) is essential for practical and fault-tolerant quantum computing (FTQC)~\cite{fowler2012surface, google2023suppressing}. Most QEC protocols impose stringent performance requirements on the physical device level due to significant hardware overhead and low error thresholds~\cite{google2023suppressing, acharya2024quantum}. The suppression of qubit error via surface code highlighted the progress in this direction~\cite{acharya2024quantum}. Alternatively, tailored QEC codes explore the noise structures of quantum systems to achieve higher efficiency and error thresholds, relaxing the requirements for the quantum device performance~\cite{michael2016new,levine2024demonstrating,aliferis2009fault, wu2022erasure}. 
QEC codes tailored for noise-biased qubits~\cite{tuckett2018ultrahigh,tuckett2019tailoring,tuckett2020fault}, which are resilient to some of the Pauli noises, such as the XZZX codes~\cite{bonilla2021xzzx,xu2022engineering,xu2023tailored}, have gained extensive attention. Recent work on enhancing the phase-flip time of dissipative cat qubits using repetition codes marks an important step toward the extreme limit where physical noises are so strongly biased that only one error channel requires correction~\cite{putterman2024hardware}. Nevertheless, the additional error channels beyond phase flips are still non-negligible~\cite{putterman2024hardware}. On the other hand, the exploration of Kerr-cat qubits (KCQs) aims to address stochastic Pauli errors based on the XZZX surface code and relies on noise bias to achieve high error thresholds~\cite{darmawan2021practical}. Importantly, resonant-KCQs have recently been shown to be promising through demonstrations of their universal gate control and quantum nondemolition readout~\cite{hajr2024high, grimm2020stabilization}.

In terms of implementation, noise bias can be achieved by encoding qubits into bosonic modes of quantum oscillators. Specifically, cat qubits~\cite{wang2016schrodinger,grimm2020stabilization,lescanne2020exponential, darmawan2021practical,gautier2022combined, bild2023schrodinger,chavez2023spectral,hillmann2023quantum, nguyen2023empowering, pan2023protecting, hajr2024high, bhandari2024symmetrically}, realized by engineering Schrödinger cat states in quantum oscillators, have attracted considerable attention due to their comparatively high noise bias and low hardware complexity. Leveraging strong non-linear interactions in superconducting circuits, many theoretical proposals examined cat qubit implementations~\cite{goto2016universal,xu2022engineering}, bias-preserving gates~\cite{guillaud2019repetition, puri2020bias}, and applications in XZZX surface code for FTQC~\cite{darmawan2021practical}. Furthermore, the successful implementation of cat qubits in superconducting circuits has driven experimental advances ~\cite{lescanne2020exponential, grimm2020stabilization}. The universal quantum operations of a KCQ under a resonant two-photon stabilization drive, or resonant-KCQ, have been demonstrated recently~\cite{hajr2024high, iyama2024observation, frattini2024observation}. In addition, a significant bit-flip time improvement has been observed on detuned-KCQs through a red-detuned two-photon stabilization drive~\cite{venkatraman2022driven}. However, noise structure characterization on KCQs has so far been limited to bit-flip and phase-flip time measurements when the qubits are idle, leading to overestimated noise bias in actual gate operations. Furthermore, while detuned-KCQs have improved bit-flip times~\cite{venkatraman2022driven}, their universal quantum operations have not yet been demonstrated. The absence of direct noise characterization for KCQ operations and universal gate implementations for detuned-KCQs has limited their further progress. Therefore, the development of a comprehensive quantum control toolbox for detuned-KCQs and a direct, accurate, and thorough characterization of the quantum gate noise structures are crucial for advancing KCQs towards further applications in QEC.

\begin{figure*}[t!]
    \centering
    \includegraphics[width=0.9\textwidth]{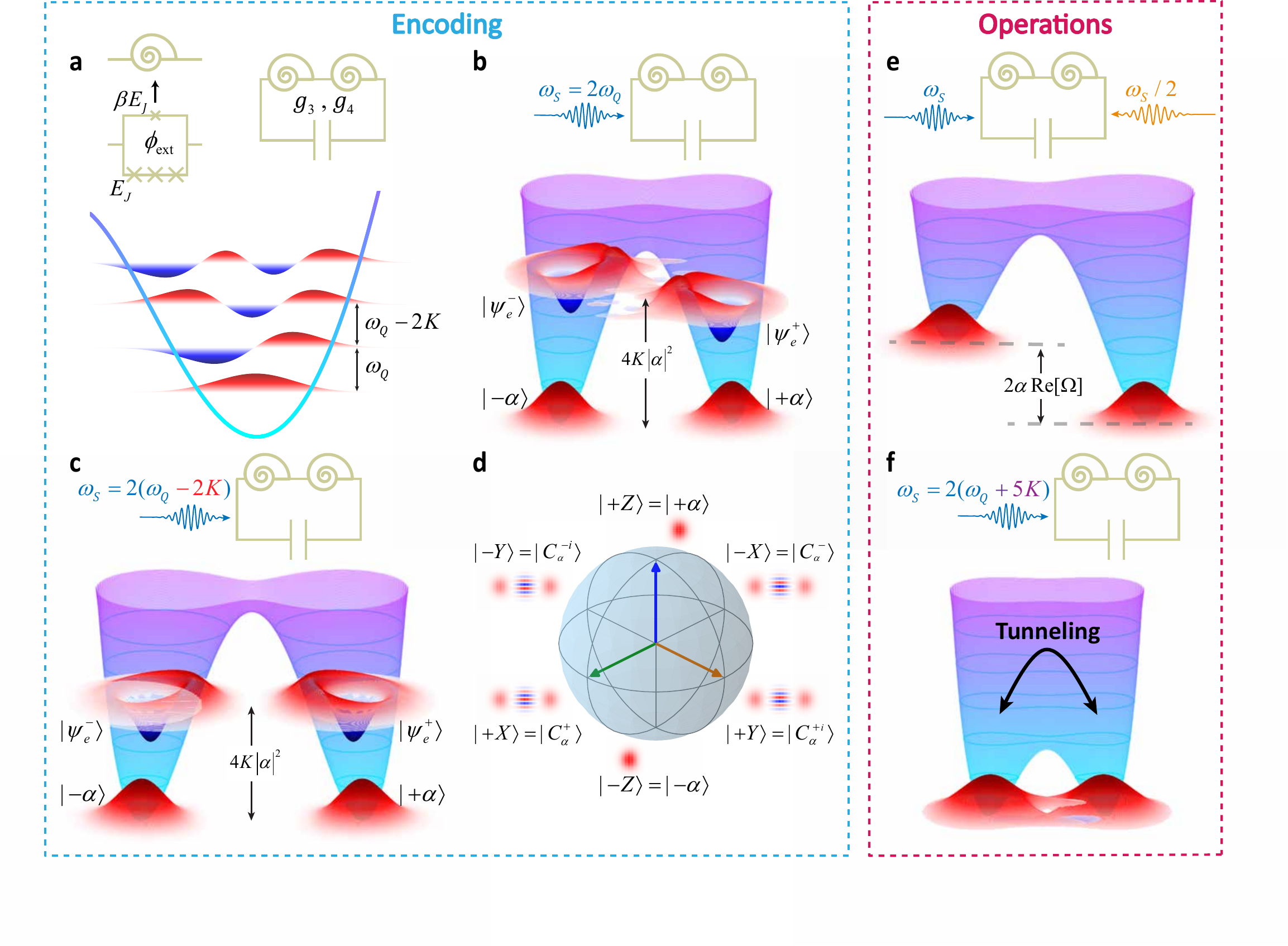}
    \caption{Kerr-cat qubit concepts. \textbf{a}, SNAIL nonlinear oscillator circuit with frequency $\omega_Q$, nonlinear factors $g_3, g_4$ and Kerr nonlinearity $K$, and its energy levels with wave-functions. \textbf{b}, SNAIL oscillator pumped by a two-photon stabilization drive at $\omega_S = 2\omega_Q$, and the engineered pseudo-potential profile with its energy levels and wave-functions. \textbf{c}, Pseudo-potential profile and its energy levels engineered by a red-detuned two-photon stabilization drive, where the excited states become degenerate. \textbf{d}, Bloch sphere defined by the coherent states and their superpositions, where $|\pm Z\rangle = |\pm \alpha\rangle$, $|\pm X\rangle = |C_\alpha^{\pm}\rangle \propto |\alpha\rangle \pm |-\alpha\rangle$, $|\pm Y\rangle = |C_\alpha^{\pm i}\rangle \propto |\alpha\rangle \pm i|-\alpha\rangle$. \textbf{e}, Pseudo-potential profile under a single-photon drive $\Omega$ at $\omega_S/2$ for $Z(\theta)$ gates. \textbf{f}, Pseudo-potential profile under a blue-detuned two-photon stabilization drive for the $X(\pi/2)$ gate. The detuning was chosen to be $5K$ as an example.}
    \label{fig:1}
\end{figure*}

In this work, we develop the first comprehensive quantum control toolbox for the detuned-KCQ in planar superconducting circuits, including the high-fidelity initialization, universal single-qubit gates and quantum nondemolition readout, with a comparison to its resonant counterpart realized in our previous work~\cite{hajr2024high}. Moreover, we directly characterize the noise structures of single-qubit operations for both the resonant- and detuned-KCQ, providing a precise and detailed description of the various noises affecting these operations. Using gate set tomography (GST), we achieved SPAM-free gate fidelities over 92.5\% for the $X(\pi/2)$ gate and 99.2\% for the bias-preserving $Z(\pi/2)$ gate, underscoring the highest performance in a noise-biased platform. Benchmarking with $\mathbb{D}_8$ dihedral group randomized benchmarking (DRB)~\cite{carignan2015characterizing, claes2023estimating} further reveals the noise bias exceeding 100 for their noise-preserving operations. Notably, the detuned-KCQ exhibits lower bit-flip errors and similar phase-flip errors compared to the resonant-KCQ, resulting in a noise bias of 250.


Interestingly, the noise bias of short gates extracted using DRB is much lower than that estimated from the bit-flip and phase-flip time measurements, signifying the inaccuracy of noise bias characterization based solely on idling lifetimes. As the first thorough and direct exploration of the noise structures of quantum operations on noise-biased qubits, we provide a framework for benchmarking the performance of quantum systems with structured noises. Moreover, thanks to the development of high-fidelity universal single-qubit operations on the detuned-KCQ, our work also positions detuned-KCQ as a strong candidate for noise-biased QEC.

\section*{Noise-Biased Cat Qubit Encoding and Control}
We extend the investigation of the resonant-KCQ introduced in our previous work~\cite{hajr2024high} to explore the detuned-KCQ by engineering the effective Hamiltonian of a superconducting nonlinear oscillator under a strong two-photon stabilization drive with various frequencies. The encoding and operations of the detuned-KCQ are illustrated in Fig.~\ref{fig:1}. The nonlinear oscillator consists of two Superconducting Nonlinear Asymmetric Inductive eLements (SNAILs) shunted by a capacitor shown in Fig.~\ref{fig:1}a. Each SNAIL consists of three large Josephson junctions with Josephson energy $E_J$ and a small Josephson junction with Josephson energy $\beta E_j$ (Suppl. Info. Section 1). For a fixed external flux $\phi_\text{ext}$, the undriven SNAIL nonlinear oscillator offers an anharmonic potential with non-zero third-order $g_3$ and fourth-order $g_4$ nonlinearities shown in Fig.~\ref{fig:1}a. 

Under a two-photon stabilization drive at frequency $\omega_S$, the effective Hamiltonian in the frame rotating at frequency $\omega_S/2$ is
\begin{equation}
    \hat{H}_\text{detuned-KCQ}/\hbar = \Delta\hat{a}^\dagger\hat{a} + \epsilon_2\hat{a}^{\dagger 2} + \epsilon_{2}^*\hat{a}^2 - K\hat{a}^{\dagger 2}\hat{a}^2,
    \label{H_KCQ}
\end{equation}
where we applied the rotating wave approximation (Materials and Methods). The non-zero detuning $\Delta = \omega_Q - \omega_S/2$ is crucial to encode the detuned-KCQ with enhanced performance. The two-photon stabilization drive $\epsilon_2$ is engineered with the third-order nonlinearity, which converts a single photon from the drive at frequency $\omega_S$ into two photons in the SNAIL oscillator at frequency $\omega_Q$, and the Kerr coefficient $K$ originates from the fourth-order nonlinearity.

This Hamiltonian forms a pseudo-potential with a double-well structure containing two degenerate ground states at the bottom of the wells, as shown in Fig.~\ref{fig:1}b for resonant-KCQ with $\Delta = 0$ and Fig.~\ref{fig:1}c for detuned-KCQ with $\Delta = 2K$. These states are separated from the excited states $|\psi_e^\pm\rangle$ by an energy gap of $4K|\alpha|^2$ with $|\alpha|^2 = (|\epsilon_2|+\Delta/2)/K$, where $|\alpha|^2$ is often used to indicate the size of the cat in phase space. These ground states approach the coherent states $|\pm\alpha\rangle$ rapidly as $|\epsilon_2|/K$ increases (Materials and Methods). By encoding these ground states as the $Z$-axis states on the Bloch sphere, their superposition spans the computational space of KCQs, as shown in Fig.~\ref{fig:1}d. 

As a noise-biased qubit, KCQ has a higher bit-flip time $T_z$ than the phase-flip time $T_y$. The coherent states on the $Z$ axis, as eigenstates of the bosonic annihilation operator, are immune to photon losses, thus leading to the increased $T_z$. However, single-photon loss flips the cat states on the equator of the Bloch sphere with a rate enhanced by $|\alpha|$, leading to the reduced $T_y$. It is important to note that, as illustrated in Fig.~\ref{fig:1}c, the excited states of the detuned-KCQ are degenerate when the two-photon stabilization drive is red-detuned by $\Delta = 2K$. Such degeneracy can lead to an even higher bit-flip time $T_z$ without further degradation of the phase-flip time $T_y$~\cite{venkatraman2022driven}, providing advantages in quantum information processing (Materials and Methods). 

\begin{figure*}[t!]
    \centering
    \includegraphics[width=1\textwidth]{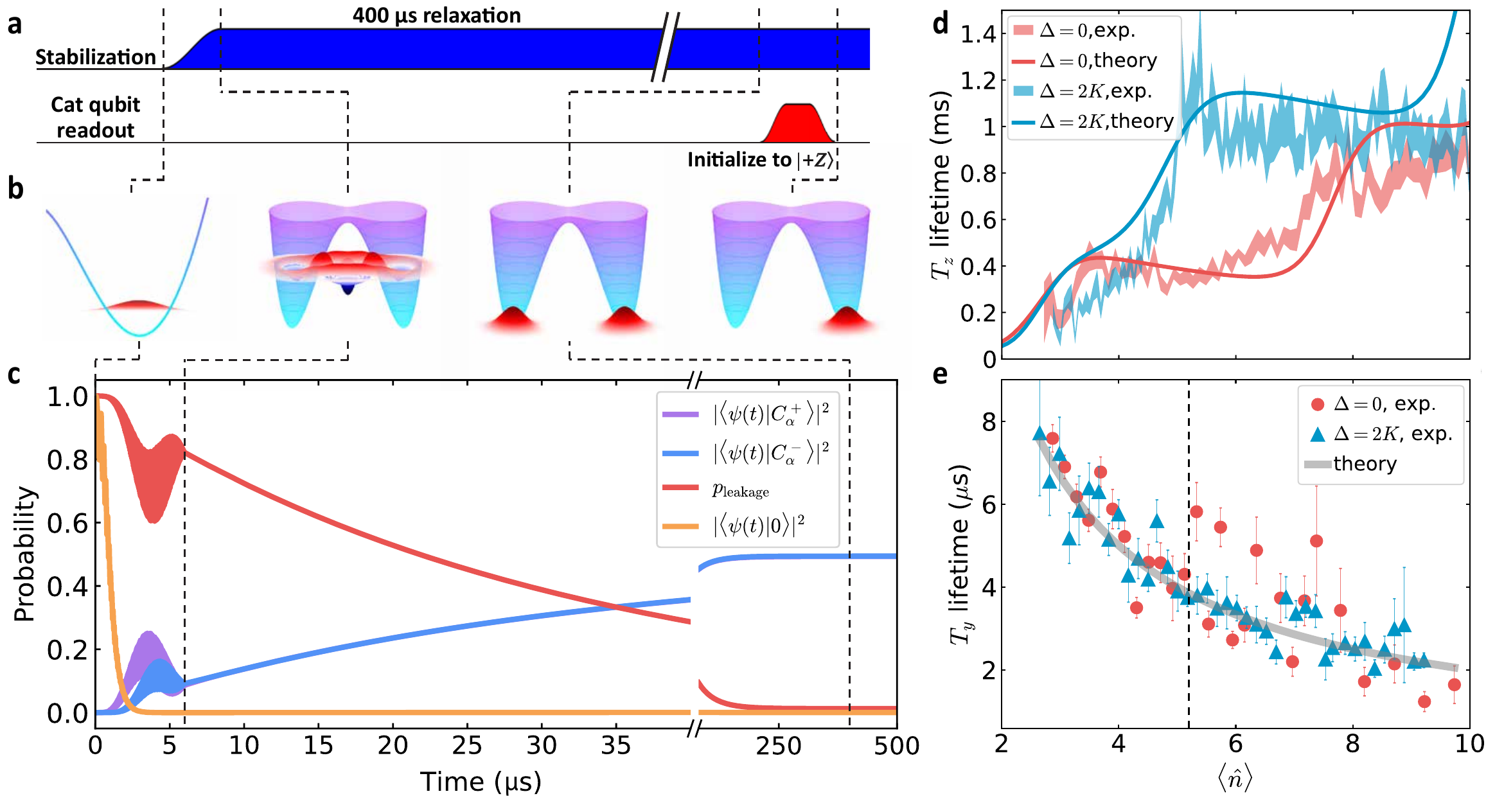}
    \caption{Initialization, bit-flip times and phase-flip times. \textbf{a}, The sequence for detuned-KCQ initialization. \textbf{b}, The evolution of the potential profiles and wave-functions during the initialization. \textbf{c}, The numerical simulation of the detuned-KCQ state population transfer during the initialization. \textbf{d}, The bit-flip time $T_z$ of resonant-KCQ and detuned-KCQ as a function of mean photon numbers. \textbf{e}, The phase-flip time $T_y$ of resonant-KCQ and detuned-KCQ. The error bars are from fitting uncertainty, and the theory predictions of the bit-flip times are calculated by master-equation simulations illustrated in ~\cite{bhandari2024symmetrically, venkatraman2022driven} and Materials and Methods.}
    \label{fig:2}
\end{figure*}

While the quantum operations of a resonant-KCQ have been extensively studied in our previous work and related literature~\cite{hajr2024high, grimm2020stabilization, frattini2024observation}, only bit-flip time measurement has been reported for detuned-KCQs with a small two-photon stabilization drive~\cite{venkatraman2022driven}, hindering its further development. Here, we demonstrate a complete quantum control toolbox for detuned-KCQ, including the initialization, universal single-qubit gates, and quantum nondemolition readout. The universal single-qubit gates consist of the continuous $Z(\theta)$ gates and a discrete $X(\pi/2)$ gate through pseudo-potential deformation. The continuous $Z(\theta)$ gates are realized by a single-photon drive $\Omega$ to lift the degeneracy between the two coherent states, allowing them to accumulate different dynamical phases, as shown in Fig.~\ref{fig:1}e. The discrete $X(\pi/2)$ gate is engineered by blue-detuning the stabilization drive to lower the energy barrier between the two coherent states, allowing them to tunnel between the two wells, as shown in Fig.~\ref{fig:1}f (Materials and Methods). In contrast to the previously reported method for resonant-KCQs based on the interruption of two-photon stabilization drive~\cite{frattini2024observation}, our method corresponds to a fidelity over 92\%, even for large cat sizes, by keeping the stabilization drive on to mitigate the effects of decays of Fock states.

\section*{Qubit Initialization, readout and Lifetime Characterization}

The KCQ is initialized to the $|+Z\rangle$ state by heralding through cat quadrature readout (CQR). CQR is realized by a beam-splitter interaction engineered through the three-wave mixing process (Materials and Methods). Under such a beam-splitter interaction, the readout resonator is populated to a steady state depending on the average value of the KCQ mode $\langle\hat{a}\rangle$. Therefore, by detecting the readout resonator state, one can determine in which well the KCQ state resides and collapse it into one of the pseudo-potential wells. This process enables the readout of KCQ along the $Z$ axis because the $|\pm Z\rangle$ states of KCQ are coherent states localized in one of the potential wells. This readout method is quantum nondemolition when the mean photon numbers in KCQs are conserved due to the much stronger two-photon stabilization drives than the beam splitter interactions. The quantum-nondemolition-ness (QNDness), defined as the probability of measuring the same results in two consecutive readouts, is observed to be over 98\%. Such high QNDness allows the initialization of qubits through heralding. 

While the resonant-KCQ can be initialized from a vacuum state by applying a CQR pulse followed immediately after ramping up the stabilization drive, a sufficiently long relaxation time before the CQR pulse is necessary to initialize the detuned-KCQ. This is because the vacuum state can be mapped to the excited states with a detuned stabilization drive. (Materials and Methods). In the experiments, the initialization sequence is shown in Fig.~\ref{fig:2}a, along with the corresponding potential profiles and wave-functions in Fig.~\ref{fig:2}b. By ramping up the two-photon stabilization drive, the potential of the SNAIL oscillator is converted to a double-well pseudo-potential. Notably, due to the detuning $\Delta$ of the two-photon stabilization drive, the vacuum state is mapped to the even-parity excited state of the pseudo-potential, which is out of the computational manifold. Therefore, a 400-$\mathcal{\mu}\text{s}$ relaxation time is added to allow this excited state to decay to the ground states in the computational manifold. Finally, the detuned-KCQ state is collapsed onto the $|+Z\rangle$ states by a CQR pulse (Materials and Methods). 

Fig.~\ref{fig:2}c shows the simulated overlaps of the detuned-KCQ state $|\psi(t)\rangle$ with the computational states $|C^\pm_\alpha\rangle = |\pm X\rangle$ and the vacuum state $|0\rangle$ during initialization with $\Delta = 2K$. The leakage out of the computational basis is defined as $p_\text{leakage}=1-|\langle\psi(t)|C^+_\alpha\rangle|^2-|\langle\psi(t)|C^-_\alpha\rangle|^2$. Another fast initialization method based on a chirp pulse has been reported recently~\cite{iyama2024observation, xu2025dynamic}, but our method achieves higher initialization fidelity despite the long relaxation times, as indicated by the high QNDness over 98\% (Materials and Methods, Fig. 7). 

We characterize the bit-flip time $T_z$ (phase-flip time $T_y$) of the KCQ by initializing it along the $Z$-axis ($Y$-axis) and measuring the decay of the Pauli operator expectation values $\langle\hat{Z}\rangle$ ($\langle\hat{Y}\rangle$). As shown in Fig.~\ref{fig:2}d and Fig.~\ref{fig:2}e, $T_z$ increases quasi-exponentially in a staircase-like manner with the mean photon number $\langle\hat{n}\rangle =|\alpha|^2$, while $T_y$ decreases according to $T_y = T_1/(2\langle\hat{n}\rangle)$~\cite{hajr2024high, jayashankar2023quantum, grimm2020stabilization, venkatraman2022static}, where $T_1\approx40~\mathrm{\mu s}$ is the relaxation time of the SNAIL oscillator. A master equation simulation, accounting for both single- and two-photon losses, photon heating and non-Markovian noise mechanisms, captured the behavior of $T_z$, with discrepancies at high $\langle\hat{n}\rangle$ primarily due to increased thermal population (Materials and Methods). 

Due to the degeneracy of higher excited states, we observed increased $T_z$ with detuning $\Delta = 2K,
4K...$ (Materials and Methods). Specifically, $T_z$ is higher for $\Delta = 2K$ compared to $\Delta = 0$ with the same mean photon number, and more importantly, the same $T_y$, as shown in Fig.~\ref{fig:2}d and \ref{fig:2}e. Therefore, a detuned-KCQ with $\Delta = 2K$ is expected to exhibit greater noise bias without additional phase-flip errors, which motivates the exploration of detuned-KCQ for enhanced performance~\cite{venkatraman2022driven,gravina2023critical}. 

In the following sections, we focus on the case where $\langle\hat{n}\rangle = 5.2$ (dashed lines in Fig.~\ref{fig:2}d and \ref{fig:2}e) and fix the detuning of the detuned-KCQ to be $2K$. At this operating point, $T_z$ is peaking for detuned-KCQ, while $T_y$ remains sufficiently high for quantum gate control. The bit-flip time $T_z$ of detuned-KCQ reaches up to $1.2~\text{ms}$, approximately three times the 0.38 ms observed for the resonant-KCQ. Meanwhile, the phase-flip time $T_y$ is around 4$~\mathrm{\mu}$s for both the resonant-KCQ and detuned-KCQ.

\section*{Gate Fidelity Characterization}

To estimate the performance of gate operations and, more importantly, the noise structures on resonant- and detuned-KCQ, we implement GST to extract the error rates, gate fidelities, and Pauli transfer matrices (PTMs) without SPAM errors (Suppl. Info. Section 5). The error channels are considered a combination of coherent and stochastic Pauli errors, described by a PTM $\mathcal{E} = e^\mathbf{L}$. The error generator $\mathbf{L}$ is defined as

\begin{equation}
\mathbf{L} = h_x \mathbf{H_x}+h_y \mathbf{H_y}+h_z \mathbf{H_z}+
p_x \mathbf{P_x}+p_y \mathbf{P_y}+p_z \mathbf{P_z},
\end{equation}
where $\mathbf{H}_x(\mathbf{P}_x), \mathbf{H}_y(\mathbf{P}_y), \mathbf{H}_z(\mathbf{P}_z)$ are the coherent (stochastic) error generators with respect to $X$, $Y$ and $Z$ axis and $h_x~(p_x), h_y~(p_y), h_z~(p_z)$ are the corresponding errors (Suppl. Info. Section 7).

\begin{figure*}[t!]
    \centering
    \includegraphics[width=1\textwidth]{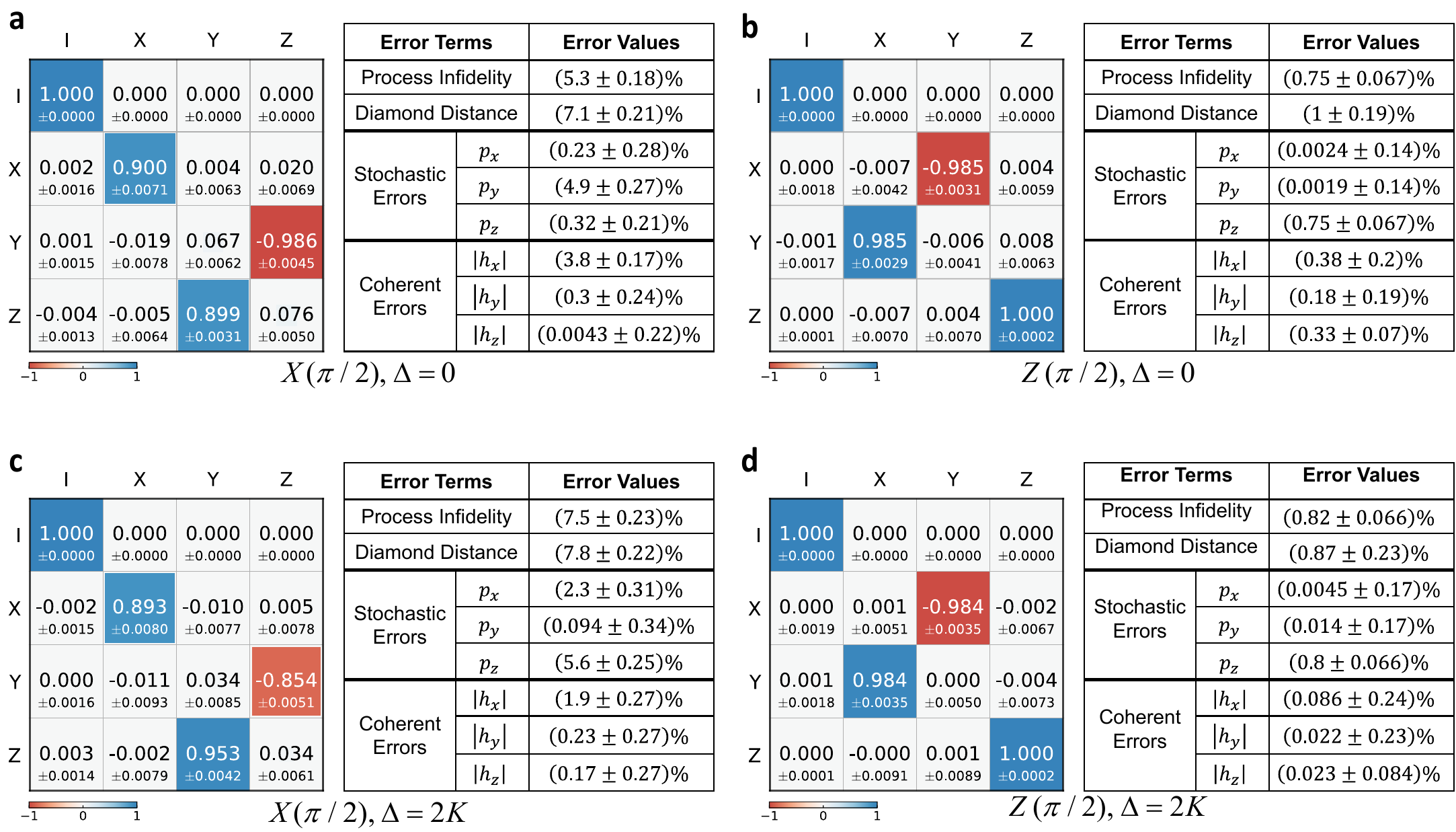}
    \caption{GST results. SPAM-free Pauli transfer matrices, process infidelities, diamond distances, stochastic Pauli errors and coherent Pauli errors for gates on resonant- and detuned-KCQs. \textbf{a}, resonant-KCQ $X(\pi/2)$ gate with 367-ns gate time. \textbf{b}, resonant-KCQ $Z(\pi/2)$ gate with 70-ns gate time. \textbf{c}, detuned-KCQ $X(\pi/2)$ gate with 418-ns gate time. \textbf{d}, detuned-KCQ $Z(\pi/2)$ gate with 70-ns gate time. The uncertainty represents the standard deviation of the model fitting, and each circuit in GST is repeated by 1024 shots.}
    \label{fig:3}
\end{figure*}

The model violation of GST under completely positive, trace preserving (CPTP)~\cite{nielsen2021gate} constraints is as small as $N_\sigma = 14.51$, indicating the validity of the completely positive and trace preserving assumption. The SPAM-free estimations of the PTMs, errors, and fidelities for $X(\pi/2)$ and $Z(\pi/2)$ gates are shown in Fig.~\ref{fig:3}. For $Z(\pi/2)$ gates, the stochastic Pauli $X$ and $Y$ errors are significantly smaller than the stochastic Pauli $Z$ errors, confirming the bias preservation. Our KCQs exhibit low coherent errors, as evidenced by similar diamond distances and process infidelity. In fact, the effects of coherent errors are negligible after Pauli twirling (Suppl. Info. Section 7). According to the extracted process infidelities of the quantum gates, the process fidelities of $Z(\pi/2)$ and $X(\pi/2)$ gates for resonant-KCQ are 99.3\% and 94.7\%, and the detuned-KCQ has similar performance with the fidelities of 99.2\% and 92.5\%, respectively, signifying state-of-the-art performance~\cite{grimm2020stabilization,hajr2024high}. The slightly lower gate fidelity of the $X(\pi/2)$ for detuned-KCQ is mainly due to the longer gate time. 

Even though GST provides a comprehensive description of the noise structures of quantum operations on KCQs and confirms the noise-bias structure for bias-preserving gates, the uncertainty of extracted stochastic Pauli $X$ and $Y$ errors for bias-preserving gates is so large that a faithful estimation of them is not possible. This is because the Pauli $X$ and $Y$ errors are too small to be measured by GST with a shallow circuit depth of 128 at most. Therefore, we will apply the noise-biased dihedral randomized benchmarking method with significantly deeper circuits to accurately extract the rare Pauli $X$ and Pauli $Y$ errors.

\section*{Noise-Bias Dihedral Randomized Benchmarking}

Quantum operations usually introduce additional noise, which may degrade the noise bias. Therefore, the naive estimations of noise-bias from bit-flip and phase-flip time measurements~\cite{venkatraman2022driven, hajr2024high, grimm2020stabilization} are not accurate, and it is essential to benchmark the noise-bias of bias-preserving operations of the KCQ. Based on the stochastic Pauli error channel, we define the bit-flip error as $p_\text{bit} = p_x+p_y$, the phase-flip error as $p_\text{ph} = p_z$, and the noise bias as $\eta = p_\text{ph}/p_\text{bit}$. The conventional randomized benchmarking method based on Clifford gates will mix the bit-flip and phase-flip errors, thereby preventing the estimation of noise bias ~\cite{knill2008randomized}. Therefore, we present the first experimental application of the DRB protocol on the KCQ to extract the noise structure of the $Z(\theta)$ gates with high precision~\cite{claes2023estimating, carignan2015characterizing} by applying a benchmarking circuit with a maximum depth over 2000. This protocol is also extensible to multi-qubit gate benchmarking by a recent protocol on CX dihedral groups~\cite{claes2023estimating}.

The noise-biased DRB protocol samples random operations from the single-qubit $\mathbb{D}_8$ dihedral group, which is generated by $X(\pi)$ and $Z(\pi/4)$ gates. The $X(\pi)$ gate for KCQ is trivially a $\pi$ phase shift of the reference, which can be applied virtually by adding an extra phase to the following control and readout pulses in a noise-free manner. This noise-free virtual $X(\pi)$ gate generates a bias-preserving $\mathbb{D}_8$ dihedral group with $Z(\pi/4)$ gates. Therefore, the DRB protocol avoids mixing the bit-flip and phase-flip errors, allowing them to be benchmarked separately. However, the noise-free property of the virtual $X(\pi)$ gate leads to an underestimation of the extracted errors of $Z(\theta)$ rotations because the extracted errors are weighted averages of both the noise-free virtual $X(\pi)$ gate and noisy $Z(\theta)$ rotations. Even though such underestimation is minor due to the low weight of the $X(\pi)$ gate in $\mathbb{D}_8$, we proceed to address this underestimation by introducing a scaling factors of 1.07 for $p_\text{bit}$ and 1.02 for $p_\text{ph}$ calculated from numerical simulations (Suppl. Info., Section 6). As a result, by applying the DRB protocol, we can precisely extract the bit-flip errors, phase-flip errors, and noise bias of the bias-preserving gates. 

\begin{figure}[t!]
    \centering
    \includegraphics[width=0.48\textwidth]{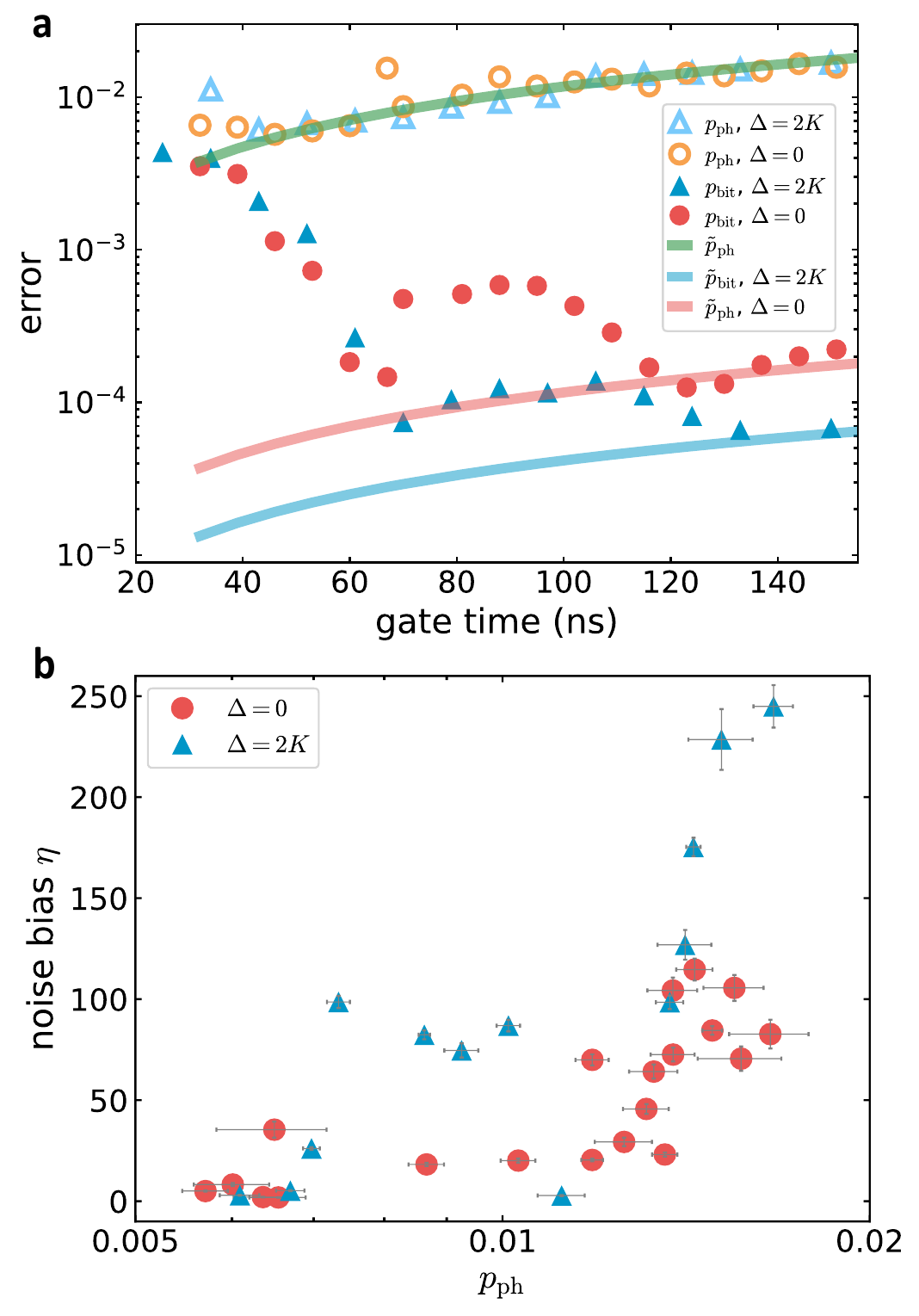}
    \caption{DRB results. \textbf{a}, The bit-flip $p_\text{bit}$ and phase-flip $p_\text{ph}$ errors of resonant- and detuned-KCQ, and corresponding estimation $\tilde{p}_\text{ph}$ and $\tilde{p}_\text{bit}$ based on the bit-flip and phase-flip times measured in Fig.~\ref{fig:2}. The error bars are smaller than the marker size and therefore not plotted. \textbf{b}, The noise bias as a function of phase-flip errors.}
    \label{fig:4}
\end{figure}

We presently explore the noise structure of the resonant-KCQ and detuned-KCQ as a function of $Z(\theta)$ gate times, with results shown in Fig.~\ref{fig:4}a, where the uncertainty of the extracted errors is significantly smaller than that from GST. We also plot the bit-flip and phase-flip errors as predicted by the bit-flip and phase-flip times based on the stochastic Pauli error model ~\cite{hashim2024practicalintroductionbenchmarkingcharacterization}. For both the resonant-KCQ and detuned-KCQ, the bit-flip error is much lower than the phase-flip error. Clearly, while the phase-flip errors align with the rates inferred from the phase-flip time (green line), the bit-flip error is typically much higher than predicted by the bit-flip time (red and blue lines for resonant- and detuned-KCQs, respectively). We attribute the high bit-flip error at short gate times to the leakage into higher excited states caused by the strong gate drive, which tunnels between the wells of the pseudo-potential. As the gate time increases, the bit-flip error approaches the error set by the bit-flip time. These observations highlight an underestimation of the bit-flip errors, and accordingly, an overestimation of the noise bias, based solely on the bit-flip and phase-flip times, underscoring the need for direct benchmarking of the operational noise characteristics.

As shown in Fig. \ref{fig:4}a, for gate times longer than 60 ns, the detuned-KCQ has a lower bit-flip error than the resonant-KCQ while maintaining the same phase-flip error, indicating the advantage of the detuned-KCQ with higher noise bias. Fig. \ref{fig:4}b further shows the noise-bias of both resonant- and detuned-KCQs with respect to their phase-flip errors. Using the detuned-KCQ, we can achieve a noise bias as high as 250 with phase-flip error rates below $2\%$ for the bias-preserving gate set. This is over twice as large as that for the resonant-KCQ without increasing the phase-flip error. Such high-fidelity bias-preserving gates with a large noise bias, together with high-fidelity quantum nondemolition readout and state initialization, represent the state-of-the-art performance and position detuned-KCQ as a promising platform for the future implementation of hardware-efficient quantum error correction.

\section*{Discussion}

In summary, we have developed a complete quantum control toolbox for detuned-KCQs, including initialization, universal single-qubit gates, and quantum nondemolition readout. Moreover, we provide the first comprehensive experimental benchmarking of noise-biased qubit operations, confirming their bias-preserving properties and revealing the inaccuracy of noise bias estimation from idling bit-flip and phase-flip times. By introducing a detuned stabilization drive to enhance the bit-flip time, we demonstrate the first implementation of high-fidelity single-qubit universal quantum operations on a detuned-KCQ, achieving a bit-flip time of 1.2 ms, which is three times the conventional resonant-KCQ, without further introducing more phase-flip noises. Using GST, we extract the detailed noise structure of the noise-biased qubit operations, and measured SPAM-free gate fidelities of over 92.5\% for the $X(\pi/2)$ gate and 99.2\% for the bias-preserving $Z(\pi/2)$ gate, representing the highest performance reported in this field. The minimal bit-flip error and the high noise bias of the bias-preserving gates are further accurately measured by the DRB protocol. While the resonant-KCQ already shows a large noise bias as high as 100 with a phase-flip error rate below 2\%, the detuned-KCQ has an even larger noise bias of 250 with a similar phase-flip error rate. With the DRB protocol, we confirmed the enhanced noise-bias without degraded phase-flip error for bias-preserving gates of detuned-KCQ, demonstrating the advantage of detuned-KCQ.

The decay rate of the SNAIL nonlinear oscillator primarily constrains the performance of the detuned-KCQ. Continued advancements in superconducting circuit materials are expected to reduce this decay rate~\cite{place2021new}, and faster gate operations enabled by pulse shaping~\cite{xu2022engineering_fast, nguyen2024programmable} and optimized circuit designs~\cite{bhandari2024symmetrically} may further enhance detuned-KCQ performance. In addition, the minor model violation observed in GST under the CPTP framework for noise characterization implies low non-trace-preserving errors, such as leakage, but a more direct and precise characterization of these errors would be beneficial for optimizing performance.

Our work exemplifies the characterization and benchmarking of noise-biased qubits, providing a framework that can be extended to other quantum systems with structured noises, such as dissipative cat qubits~\cite{lescanne2020exponential, putterman2024hardware}, GKP qubits~\cite{hanggli2020enhanced, grimsmo2021quantum, lachance2024autonomous}, and binomial qubits~\cite{ni2023beating}. Compared to other bosonic qubit platforms with biased noise, the detuned-KCQ uniquely achieves high-fidelity quantum nondemolition readout and universal gate operations simultaneously with low hardware complexity. It is also compatible with recently demonstrated two-qubit gate operations in experiment~\cite{hoshi2025entangling}. Leveraging a scalable planar superconducting architecture, our work is well-positioned for extension into multi-qubit processors, paving the way for the application of noise-biased qubits in QEC. Moreover, the enhanced bit-flip time of the detuned-KCQ presents it as a promising ancilla for universal control of a quantum cavity, effectively mitigating parasitic error propagation~\cite{puri2019stabilized, ding2024quantum}. As a Kerr parametric oscillator (KPO) with the double-well potential, the detuned-KCQ also opens new possibilities for KPO-network-based Ising machines, fostering advancements in analog quantum computing~\cite{alvarez2024biased,goto2018boltzmann,yamaji2023correlated}. 

\section*{Materials and methods}

\subsection*{Detuned-KCQ Hamiltonian and Eigenstates}

The detuned-KCQ Hamiltonian is engineered through a superconducting nonlinear resonator under a strong two-photon stabilization drive. The SNAILs in the resonator provide the nonlinear potential. 

\begin{equation}
    U_{\text{SNAIL}}(\phi) = -\beta E_J \cos\phi - 3 E_J \cos\left(\frac{\phi_\text{ext}-\phi}{3}\right),
    \label{U_SNAIL}
\end{equation}
where $E_J/\hbar = 2\pi\times263.2~\mathrm{GHz}$ is the Josephson energy of the large junction in SNAILs, $\beta = 0.1$ is Josephson energy ratio between the small and large junctions, $\phi_{\mathrm{ext}}$ is the scaled external magnetic flux threading the SNAIL loop, and $\phi$ is the associated superconducting phase across each SNAIL.

For a fixed external flux, by Taylor expanding the potential energy up to the fourth order, 

\begin{equation}
    U_{\text{SNAIL}}(\phi) \approx U_{\text{SNAIL}}(\phi_{{\text{min}}})+\sum_{k=2}^4 g_k(\phi-\phi_{\text{min}})^k,
    \label{U_expansion}
\end{equation}
one can observe that the SNAIL offers both non-zero third-order and fourth-order nonlinearities necessary for engineering the detuned-KCQ Hamiltonian. Under a two-photon stabilization drive, applying the rotating wave approximation in the rotating frame yields the effective Hamiltonian $\hat{H}_\text{detuned-KCQ}$ shown in the main text. Without loss of generality, the two-photon stabilization drive will be assumed to be in-phase, i.e. $\epsilon_2\in \mathbb{R}$.

To analyze the eigenenergy and eigenstates of this Hamiltonian, a displacement transformation $\hat{D}(\alpha)$ is introduced to obtain the corresponding Hamiltonian in the displacement frame written as 

\begin{subequations}
\begin{align}
\hat{H}_{\text{detuned-KCQ}_\text{disp}}=& E + \Lambda a^\dagger + \Lambda^* a + \tilde{\Delta}a^\dagger a + \tilde{\epsilon}_2 (a^{\dagger2} +  a^2) \\&+ \Gamma a^{\dagger2}a + \Gamma^*a^\dagger a^2 - Ka^{\dagger2}a^2,
\end{align}
\end{subequations}
where the coefficients are given by
\begin{subequations}
    \begin{align}
        E &= \Delta|\alpha|^2-K|\alpha|^4+\epsilon_2(\alpha^2+\alpha^{*2}),
        \\
        \Lambda &= -\Delta\alpha+2K|\alpha|^2\alpha^* - 2\epsilon_2\alpha^*,
        \\
        \tilde{\Delta} &= \Delta-4K|\alpha|^2,
        \\
        \tilde{\epsilon}_2 &= -K\alpha^2+\epsilon_2,
        \\
        \Gamma &= -2K\alpha.
    \label{coeff}\end{align}
\end{subequations}

With a suitable choice of $\alpha = \sqrt{(\epsilon_2+\Delta/2)/K}$, the single-photon drive terms are canceled out (i.e. $\Lambda = 0$), and the displaced Hamiltonian can be written as

\begin{subequations}
    \begin{align}
        \hat{H}_{\text{detund-KCQ}_\text{disp}}=&H_0 + H_1,\\
        H_0 =& \frac{\left(\Delta/2 + \epsilon_2\right)^2}{K}- \left(4\epsilon_2+\Delta\right)a^\dagger a,\\
        H_1 =&  -\frac{1}{2}\Delta (a^{\dagger2} +  a^2) \\&-(2\epsilon_2+\Delta)( a^{\dagger2}a + a^\dagger a^2) - Ka^{\dagger2}a^2,
        \label{perturb_H}
    \end{align}
\end{subequations}
where the eigenstates and energies of $H_0$ are Fock states $|n\rangle$ with energy $E_n^{(0)} = \left(\Delta/2 + \epsilon_2\right)^2/K- \left(4\epsilon_2+\Delta\right)n$, and $H_1$ is treated as a perturbation as long as its energy scale is much smaller than the energy gap of the unperturbed system ($|\epsilon_2|\gg|K|, |\Delta/2|$).

Firstly, the perturbed ground state energy is estimated as
\begin{equation}
    \tilde{E}_0 = E_0^{(0)} + E_0^{(1)}+\ldots = \frac{(\Delta/2+\epsilon_2)^2}{K}+\frac{\Delta^2K}{(4\epsilon_2+\Delta)^2}+\ldots,
    \label{perturb_E0}
\end{equation}

\begin{figure}[t!]
    \centering
    \includegraphics[width=0.48\textwidth]{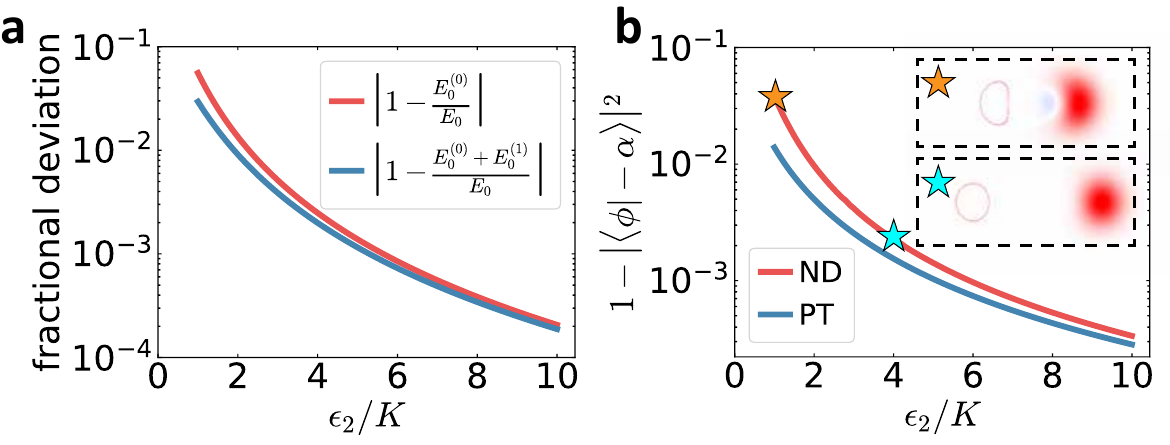}
    \caption{Detuned-KCQ ground state and its energy with $\Delta = 2K$. \textbf{a}, The fractional deviation of the ground state energy estimated by the unperturbed Hamiltonian and perturbation theory. \textbf{b}, The difference between the coherent state and the ground state of the detuned-KCQ Hamiltonian calculated by numerical diagonalization (ND) and perturbation theory (PT).}
    \label{fig:supp_energy}
\end{figure}

In Fig.~\ref{fig:supp_energy}a, we plot the fractional deviation of the estimated ground state energy $(\tilde{E_0})$ from the exact ground state energy $(E_0)$ given by the numerical diagonalization with various two-photon stabilization drive strengths. In the regime of interest where $\Delta/K \sim 2$ and $\epsilon_2/K\sim 1 - 10$, the unperturbed ground state energy $E_0^{(0)}$ matches the exact ground state energy with less than 5\% deviation, which is further improved slightly by the perturbation theory.

Secondly, the perturbed ground state (up to the first order) is given by

\begin{equation}
        |\psi_0\rangle = \mathcal{N}\left(|0\rangle - \frac{\sqrt{2}\Delta}{4(4\epsilon_2+\Delta)}|2\rangle\right),\label{perturb_psi0}
\end{equation}
where the perturbation introduces the occupation beyond the unperturbed ground states, and $\mathcal{N}$ is the normalization factor. After returning to the undisplaced frame, the ground state ends up approximately as the coherent state,
\begin{equation}
    |\phi\rangle = \hat{D}^\dagger(\alpha)|\psi_0\rangle = \mathcal{N}\left(|-\alpha\rangle - \frac{\sqrt{2}\Delta}{4(4\epsilon_2+\Delta)}\hat{D}^\dagger(\alpha)|2\rangle\right)\approx |-\alpha\rangle.\label{perturb_psi0_lab}
\end{equation}

The difference between the coherent state $|-\alpha\rangle$ and the ground state $|\phi\rangle$ of the detuned-KCQ when $\Delta = 2K$ is shown in Fig.~\ref{fig:supp_energy}b. The Wigner functions of the ground states when $\epsilon_2/K = 1$ and $\epsilon_2/K = 4$ are also plotted in the inset indicated by the orange and blue stars. Both results from perturbation theory and numerical diagonalization indicate the rapidly decreasing difference between the coherent state and the detuned-KCQ ground state. At the operating point in this work, such a difference is below 0.3\%.

\subsection*{Detuned-KCQ Bit-flip Time $T_z$ Simulations}

We illustrate the simulation of bit-flip times $T_z$ in this section and emphasize the enhancement of bit-flip times when the two-photon stabilization drive is detuned by $2K, 4K...$ due to the degenerate higher excited states~\cite{venkatraman2022driven, frattini2024observation, hajr2024high}. 

\begin{figure*}[t]
    \centering
    \includegraphics[width=1.02\textwidth]{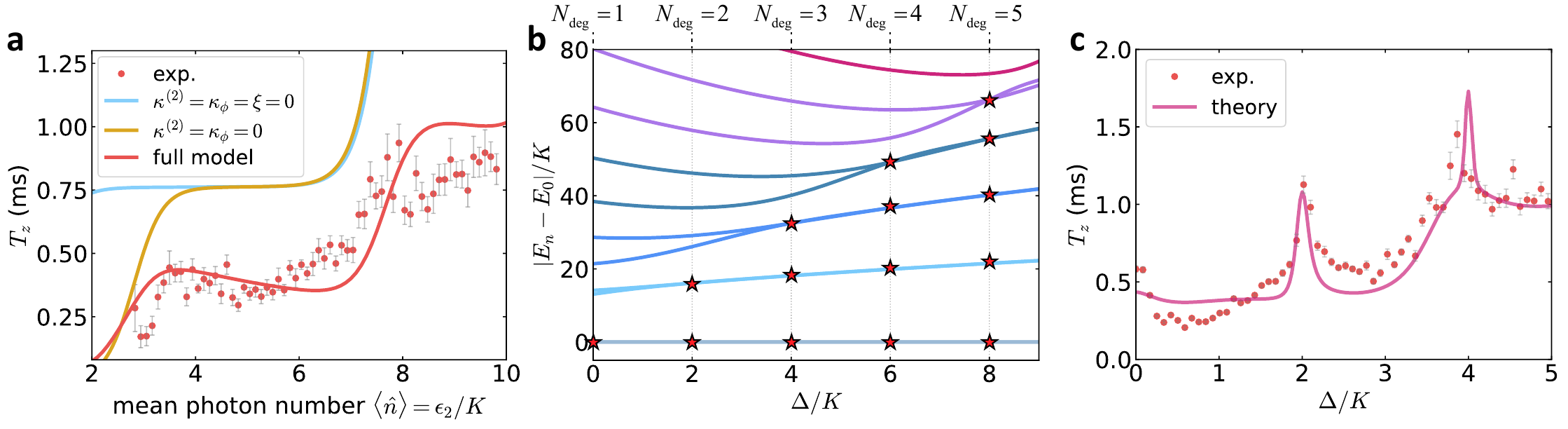}
    \caption{Master equation simulation. \textbf{a}, The bit-flip times $T_z$ with different mean photon number and their master equation simulation with the fitting parameters taking different incoherent processes into account. \textbf{b}, detuned-KCQ Hamiltonian spectrum with detuned two-photon stabilization drive. \textbf{c}, The bit-flip times $T_z$ with different detuning and their master equation simulation.}
    \label{fig:supp_4}
\end{figure*}

The Kerr-cat qubit inevitably disperses irretrievable information to the environment, leading to incoherent processes. We model the environment by a chain of harmonic oscillators with infinite modes and temperature $T_{\rm E}$. While the exact dynamics of arbitrary system-environment interaction and environment evolution are intractable analytically, one can construct the master equation describing various effects of the environment on the system under assumptions that generally hold in the experiments, including weak system-environment coupling and a Markovian thermal environment. For simplicity, we consider an environment with fixed thermal distribution and energy-independent coupling with the system. With the above approximations, the master equation has the following form,
\begin{equation}
\frac{\text{d}\hat{\rho}}{\text{d}t}=-i[\hat{H},\hat{\rho}] + \sum_l \gamma_l {\mathcal{D}}[\hat{\cal O}_l]\hat{\rho},
\end{equation}
where $\hat{\cal O}_l$ describe three sets of incoherent processes, including $\hat{\cal O}_{1,\downarrow}~(\hat{\cal O}_{1,\uparrow})= \hat{a}~(\hat{a}^\dagger)$ for the single-photon dissipation (excitation) process, $\hat{\cal O}_{2,\downarrow}~(\hat{\cal O}_{2,\uparrow})= \hat{a}^2~(\hat{a}^{\dagger^2})$ for the two-photon dissipation (excitation) process, and $\hat{\cal O}_\phi = \hat{a}^\dagger \hat{a}$ for the dephasing process. The corresponding coefficients $\gamma_{1,\downarrow}(\gamma_{1,\uparrow}) \propto \kappa^{(1)}$, $\gamma_{2,\downarrow}(\gamma_{2,\uparrow}) \propto \kappa^{(2)}$, and $\gamma_\phi \propto \kappa_\phi$ are single-photon dissipation (excitation), two-photon dissipation (excitation), and dephasing rates, where $\kappa^{(1)}$ and $\kappa^{(2)}$ are the single-photon and two-photon spontaneous decay rates, respectively. A detailed derivation of the master equation along with the expressions for $\gamma_{1,\downarrow}(\gamma_{1,\uparrow}), \gamma_{2,\downarrow}(\gamma_{2,\uparrow})$ and $\gamma_\phi$ is presented in references~\cite{venkatraman2022static,hajr2024high, bhandari2024symmetrically}. Beyond the three incoherent processes above, we also consider a stochastic process that leads to fluctuations in the SNAIL nonlinear oscillator frequency with a noise strength of $\xi$~\cite{hajr2024high, bhandari2024symmetrically}. 

\begin{figure*}[t!]
    \centering
    \includegraphics[width=0.97\textwidth]{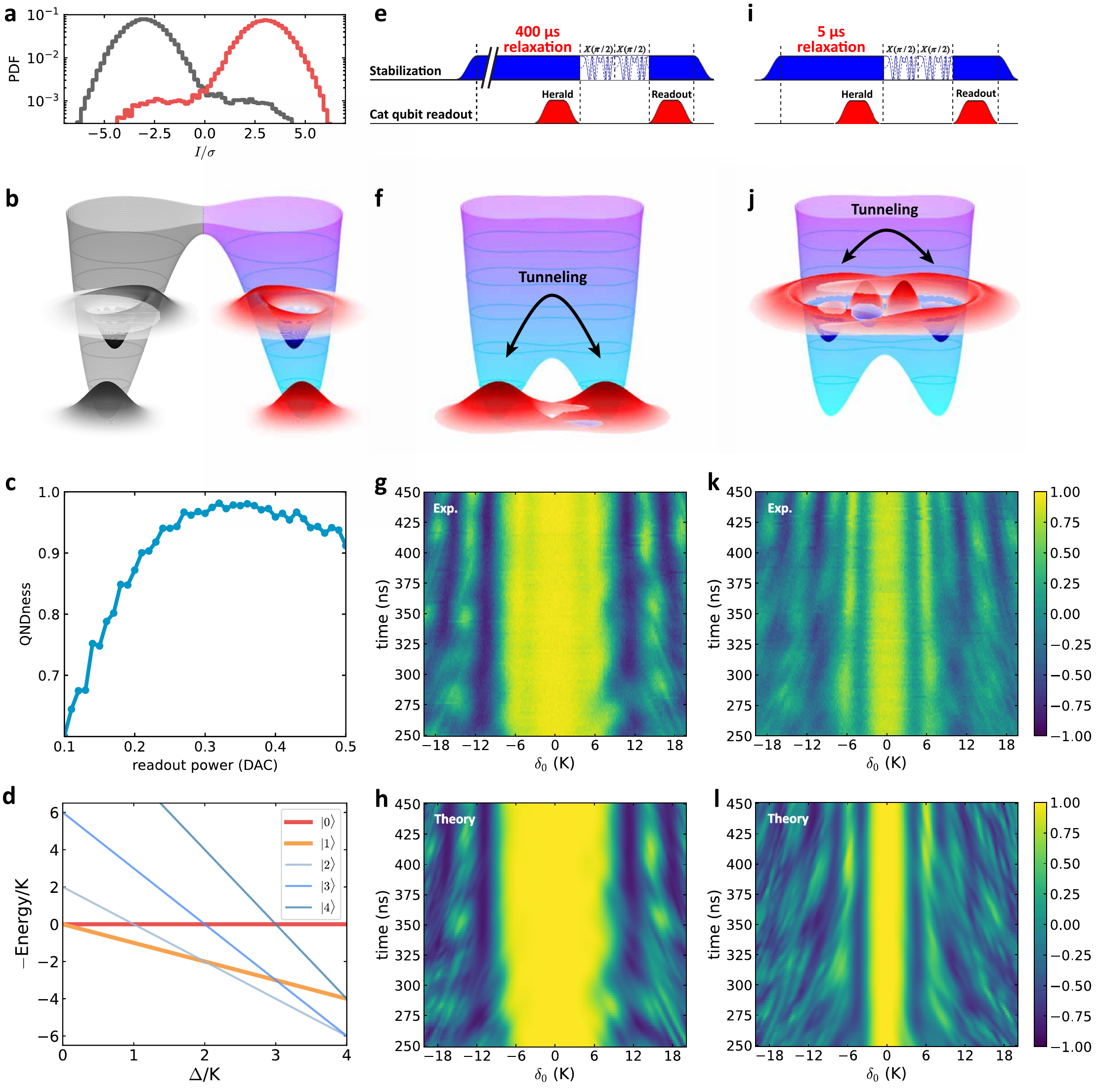}
    \caption{Detuned-KCQ readout, initialization and gate calibration. \textbf{a}, The histogram of CQR. \textbf{b}, The collapse of wave-functions corresponding to CQR. \textbf{c}, The QNDness optimization of CQR with various readout strengths. \textbf{d}, The SNAIL nonlinear oscillator spectrum in the rotating frame as a function of detuning. \textbf{e}, The sequence for calibrating $X(\pi/2)$ gate with 400-$\mathbf{\mu}s$ relaxation time in the initialization. \textbf{f}, The tunneling between the ground states when the relaxation time is long enough in the initialization. \textbf{g}, The measured $|\pm Z\rangle$ population oscillation. \textbf{h}, The numerical simulation of the results above it. \textbf{i}, The sequence for observing the excited states tunneling with 4-$\mathbf{\mu}s$ relaxation time in initialization. \textbf{j}, The tunneling between the excited states when the relaxation time is short. \textbf{k} The measured population oscillation between excited states. \textbf{l}, The numerical simulation of the results above.}
    \label{fig:7}
\end{figure*}

With such four processes in a thermal environment with thermal photon population $n_\text{th} = 4\%$, and parameters $\kappa^{(2)}=7\text{MHz}$, $\kappa_\phi=100\text{Hz}$, $\xi=40\text{kHz}$, the experimental behavior of the bit-flip times $T_z$ is accurately modeled as shown by the agreement between the experiment data and master equation simulation results in Fig.~\ref{fig:2}. The thermal population corresponds to a temperature of 86.9 mK, which is higher than the base temperature of the dilution fridge due to the strong microwave drives. The pure dephasing rate $\kappa_\phi$ is lower than that estimated from the $T_2$ values, which is also observed in all related literature~\cite{gautier2022combined, frattini2024observation, bhandari2024symmetrically}, and might be due to the non-uniform power spectrum of the dephasing process under strong drive.

Here, we illustrate the effects of each process in detail. As shown in Fig.~\ref{fig:supp_4}a, we fit the bit-flip times of a resonant-KCQ as a function of two-photon stabilization drive strength $\epsilon_2$. When only considering the single-photon dissipation and excitation, i.e. $\kappa^{(2)}=\kappa_\phi = \xi = 0$, the master equation simulation follows the blue curve, overestimating the bit-flip times $T_z$. By introducing the stochastic frequency fluctuation of the SNAIL nonlinear, i.e., $\kappa^{(2)}=\kappa_\phi = 0$, the master equation simulation successfully modeled the bit-flip times with small $\epsilon_2$ following the yellow curve, but failed to model the strongly driven regime. Finally, only by including all four processes does the master equation simulation accurately model the experimental data, as shown by the red curve.

Beyond the influence on bit-flip times $T_z$ by the two-photon stabilization drive strength, the detuning of two-photon stabilization drives, with $\Delta = 2K, 4K...$, can lead to degenerate higher excited states marked by the stars in Fig.~\ref{fig:supp_4}b, and increase the bit-flip times. We observe $N_\text{deg}$ degenerate points, corresponding to $N_\text{deg} - 1$ higher excited degenerate states, when $\Delta = 2(N_\text{deg} - 1)K$. With a fixed two-photon stabilization drive strength $\epsilon_2 = 4.2K$, we measured the bit-flip times with various detuning $\Delta$, as shown in Fig.~\ref{fig:supp_4}c. The increase of the bit-flip times when $\Delta = 2K$ and $4K$ is observed both in experimental data and predicted in the master equation simulations, demonstrating the validity of the theoretical model.

\subsection*{Quantum control Toolbox for a Detuned-KCQ}

In order to operate a system as a qubit for quantum computing, it is crucial to have reliable state readout, universal quantum control and state initialization. 

The readout of the detuned-KCQ is enabled by a beam-splitter interaction between the detuned-KCQ mode $\hat{a}$ and the readout resonator mode $\hat{b}$. Due to the detuning of the stabilization drive, the rotating frame frequency is referenced at half of the stabilization drive frequency instead of the qubit frequency. Therefore, by applying a cat-qubit-readout (CQR) drive at frequency $\omega_{CQR} = \omega_R - \omega_S/2$, we can engineer a beam-splitter interaction given as

\begin{equation}
    \hat{H}_{BS} = \epsilon_{CQR}a^\dagger b + \epsilon_{CQR}^*a b^\dagger,
    \label{BS_inter}
\end{equation}
where $\omega_R$ is the readout resonator frequency and $\epsilon_{CQR}$ is the interaction rate.

Such an interaction will populate the readout resonator mode $\hat{b}$ depending on the expectation value of the detuned-KCQ mode $\hat{a}$ as
\begin{equation}
    \langle \hat{b}\rangle(t\rightarrow+\infty) = \frac{2\epsilon_{CQR}}{\kappa_R}\langle \hat{a}\rangle,
    \label{CQR}
\end{equation}

To calibrate the CQR drive, we apply two consecutive CQR pulses with a duration of 4.5 $\mathbf{\mu}s$ for each, and define the QNDness as $(P\left[M_{|+\alpha\rangle}|M_{|+\alpha\rangle}\right] + P\left[M_{|-\alpha\rangle}|M_{|-\alpha\rangle}\right])/2$, where $P\left[M_{A}|M_{B}\right]$ represents the probability of getting result $A$ with the second CQR pulse conditioned on having obtained result $B$ with the first CQR pulse. The QNDness is optimized with respect to the CQR drive strengths shown in Fig.~\ref{fig:7}c, resulting in good separations shown by the histogram in Fig.~\ref{fig:7}a. Such a quantum non-demolition readout can be utilized to initialize the qubit through heralding. However, as shown in Fig.~\ref{fig:7}b and discussed later in this section, the qubit states have to stay in the ground states before heralding because such readout cannot distinguish the excited states and the ground state in the same potential well.

The universal single-qubit control of the detuned-KCQ consists of a discrete $X(\pi/2)$ gate and continuous $Z(\theta)$ rotation gates, both implemented via deformation of the double-well pseudo-potential. The $Z(\theta)$ gates are realized using a single-photon drive at frequency $\omega_S/2$ to engineer a Hamiltonian $\hat{H}_d = \hat{a}^\dagger\Omega/2 + \hat{a}\Omega^*/2$. This Hamiltonian induces an energy splitting of $2\text{Re}[\Omega]\alpha$ between the $|\pm Z\rangle$ states, resulting in Rabi oscillations on the equator of the Bloch sphere. Calibration and numerical simulations of the $Z(\theta)$ gates are detailed in Suppl. Info. Section 3. The $X(\pi/2)$ gate, which corresponds to coherent tunneling between the two pseudo-potential wells, is implemented by adiabatically lowering the energy barrier separating them. We implement such gate by phase-modulating the two-photon stabilization drive $\epsilon_2(t) = \epsilon_2(0)e^{-ig(t)}$ with $g(t)$ defined as

\begin{equation}
   g(t) =\delta_0t\times 
\left\{
	\begin{array}{ll}
		- \sin\left(\frac{3\pi}{2} t/T_g\right)  &  t \leq T_g/3 \\
		-\frac{f(t)}{1-f(T_g)} \left(f(t)-f(T_g)\right) &  t > T_g/3
	\end{array},
\right.
\label{modulation}
\end{equation}
where $T_g$ is the gate time, $\delta_0$ is the modulation depth, and $f(t)=\text{exp}\left(-\frac{8(t-Tg/3)^2}{T_g^2}\right)$ follows a Gaussian profile. The asymmetric modulation profile is chosen to minimize the gate error. Such phase modulation adiabatically introduces a blue-detuning to the two-photon stabilization drive and correspondingly, lowers the energy barrier of the pseudo-potential to allow the inter-well tunneling, as shown in Fig.~\ref{fig:7}f.

We calibrate the $X(\pi/2)$ gate with the sequence shown in Fig.~\ref{fig:7}e. A CQR pulse (labeled ``Herald'') initializes the qubit into $|+Z\rangle$ state through heralding. Then, two $X(\pi/2)$ gate pulses with various modulation depth $\delta_0$ and duration are applied. Finally, another CQR pulse (labeled ``Readout'') is applied to read out the qubit state. The measured expectation of the Pauli Z operator $\langle\hat{Z}\rangle$ is shown in Fig.~\ref{fig:7}g, which clearly shows population oscillating coherently between the $|\pm Z\rangle$ states. The numerical simulation, based on solving the time-dependent Schrödinger equations, is shown in Fig.~\ref{fig:7}h in good agreement with the experimental results. Therefore, we calibrate the $X(\pi/2)$ gate for a detuned-KCQ by optimizing the gate time and the modulation depth to achieve the largest $\langle\hat{Z}\rangle$ conversion. Importantly, a sufficiently long relaxation time (400 $ \mathrm{\mu s}$) is necessary for the initialization as discussed in the following.

The initialization of the detuned-KCQ is implemented through the CQR pulse by heralding. The SNAIL nonlinear oscillator mode predominantly remains in the vacuum state $|0\rangle$ with negligible thermal noise at low temperature. Therefore, the resonant-KCQ can be initialized by adiabatically ramping up the two-photon stabilization drive followed immediately by a CQR drive, which maps the vacuum state to the cat state and then collapses it into $|\pm Z\rangle$. However, this technique fails for the detuned-KCQ because the finite detuning changes the energies of SNAIL nonlinear oscillator states in the rotating frame. As shown in Fig.~\ref{fig:7}d, a finite $\Delta$ alters the spectrum in the rotating frame, causing $|0\rangle$ to no longer be the ground state, while the cat states remain the ground states of the detuned-KCQ Hamiltonian. Consequently, when the two-photon stabilization drive is ramped up, the $|0\rangle$ state is mapped to the higher excited states instead of the ground states. The subsequent CQR drive will then collapse the qubit into excited states in one of the double wells of the pseudo-potential, resulting in a failure to initialize the qubit into $|\pm Z\rangle$ states. To address this, it is crucial to introduce sufficient relaxation time before applying the CQR drive, allowing the system to decay into the degenerate ground states. This ensures the validity of preparing the $|\pm Z\rangle = |\pm\alpha\rangle$ state through heralding. 

To observe the effect of higher excited states, we applied a sequence similar to that used for calibrating the $X(\pi/2)$ gate to a detuned-KCQ system with $\Delta = 2K$, but with the relaxation time reduced to $5~\mathcal{\mu}s$, as shown in Fig.~\ref{fig:7}i. Due to this shortened relaxation time, the detuned-KCQ states predominantly occupy the excited states, and the pulse sequence captures the population oscillation between these excited states within each well, as illustrated in Fig.~\ref{fig:7}j. The experimental results, presented in Fig.~\ref{fig:7}k, show a reduced contrast and tunneling occurring at a lower modulation depth due to the low lifetime and high energy of the excited states, which is in excellent agreement with the numerical simulation shown in Fig.~\ref{fig:7}l.

\acknow{We are grateful to Chuanhong Liu, Ravi Naik and Trevor Chistolini for fruitful discussions on experiments. We thank Shruti Puri and Jahan Claes for insightful suggestions.

This work was supported by the U. S. Army Research Office under grant
W911NF-22-1-0258.

B.Q., Ah.H. and K.W. led and organized the project. B.Q. fabricated the device, performed the measurement and analyzed the data. Ah.H. and G.K. designed the device. Ah.H, G.K., and L.B.N. set up the cryogenic apparatus and initialized the measurement. I.H. and B.B. performed the theoretical analysis. Ah.H., K.W., N.J., N.G. and N.F. assisted with the measurement. J.H. offered the GST circuit and analyzed the GST data. Ak.H. assisted with the GST data analysis. Z.P., L.C., Z.K., C.J., H.K., and K.L. assisted with the device fabrication. J.D., A.J., D.S. and I.S. supervised the project and offered necessary supports. B.Q., K.W. and L.B.N. wrote the manuscript with input from all authors. }

\showacknow{} 
\bibliography{pnas-sample}

\onecolumn
{\Huge \centering Supplementary Information}

\section{Measurement Hardware, Device Layout and Device Parameters}

\begin{figure}[h!]
    \centering
    \includegraphics[width=1\textwidth]{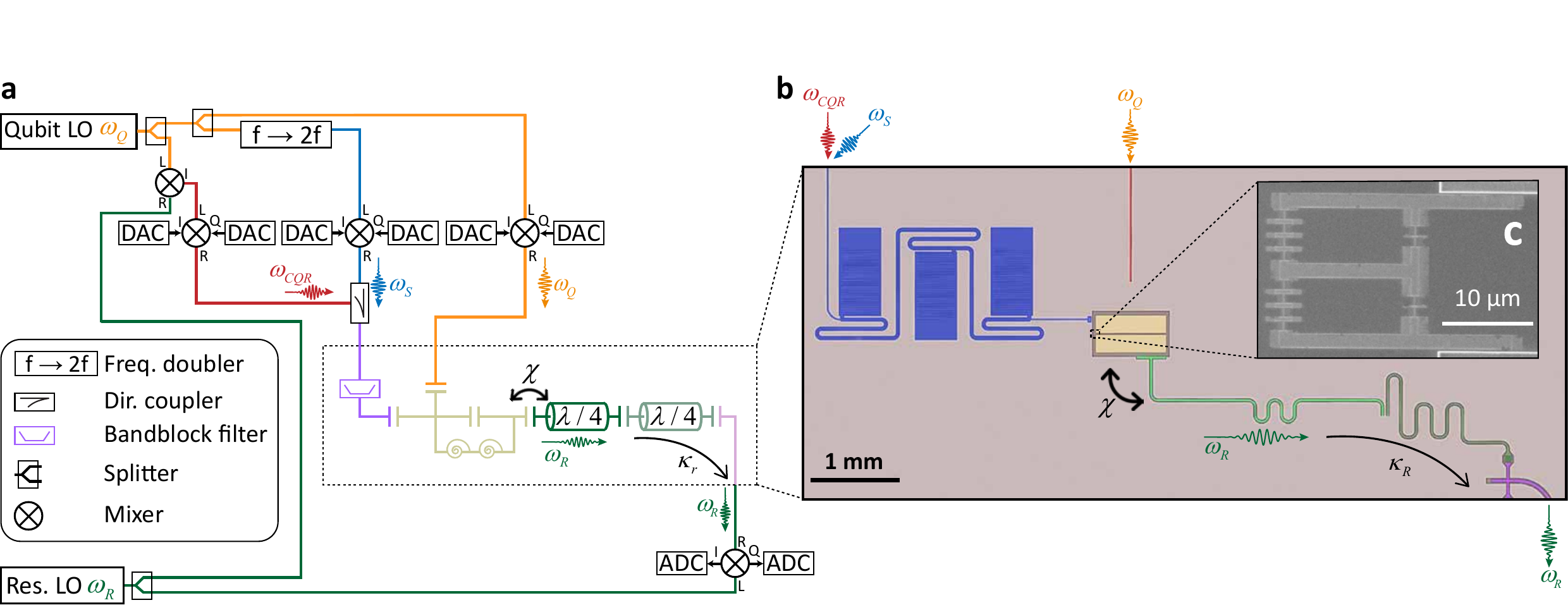}
    \caption{Device and hardware. \textbf{a} Simplified wiring diagram of the chip and the measurement hardware, where the digital-to-analog converters (DACs) are implemented by an arbitrary waveform generator (AWG), and the analog-to-digital converters (ADCs) are implemented by an Alazar sampling card. \textbf{b} False-color micrograph of the Kerr-cat qubit device, where the SNAIL nonlinear resonator (yellow) is coupled to a readout resonator (dark green) with its Purcell filter (light green) for quantum nondemolition readout. A weakly coupled control port (orange) delivers the single-photon drive, and a strongly coupled pump port with a bandblock filter (purple) delivers the two-photon stabilization drive and cat-qubit readout drive with suppressed Purcell decay. \textbf{c} The scanning electron microscope picture of the two SNAILs.}
    \label{fig:Chip}
\end{figure}

The Kerr-cat qubit is implemented on a planar superconducting circuit device. The equivalent circuit diagram of the device with the measurement wiring diagram is shown in Fig. \ref{fig:Chip}a, and the false-color micrograph of the device chip is shown in Fig. \ref{fig:Chip}b. The key component of the Kerr-cat qubit device, SNAIL nonlinear oscillator (yellow), consists of two SNAILs shunted by a coplanar capacitor, where each SNAIL is formed by three large Josephson junctions connected in parallel with a small Josephson junction, as shown in Fig, \ref{fig:Chip}c. A readout resonator (dark green) is capacitively coupled to the SNAIL nonlinear oscillator with a dispersive shift $\chi$ to enable the quantum nondemolition readout of the Kerr-cat qubit, and a Purcell filter (light green) is introduced between the readout resonator and the readout port (pink) to engineer the desired readout resonator decay rate $\kappa_r$ with suppressed Purcell decay. We engineered two microwave ports to deliver the microwave efficiently without introducing too much Purcell decay. On the one hand, a control port (orange) is weakly coupled to the SNAIL nonlinear oscillator to deliver the single-photon drive for the continuous $Z(\theta)$ gates. On the
other hand, a pump port (purple) with a bandblock filter is strongly coupled to the SNAIL nonlinear oscillator to deliver the two-photon stabilization drive and the cat-qubit readout (CQR) drive, where the bandblock filter strongly suppressed the Purcell decay from this port by blocking the signal around the qubit frequency. Detailed description of the bandblock filter can be found in our previous work\cite{hajr2024high}.

We characterize the Kerr-cat qubit with the microwave network shown in Fig. \ref{fig:Chip}a, where we neglect some filters, amplifiers and attenuators for simplicity and clarity, and a detailed wiring diagram can be found in our previous work\cite{hajr2024high}. The modulation and demodulation of the microwave drive is implemented by two local oscillators, frequency mixers, and a frequency doubler. A readout local oscillator (LO) at readout resonator frequency $\omega_R$ is introduced to demodulate the readout signal from the readout port (pink) of the device, and a qubit LO at SNAIL nonlinear oscillator frequency $\omega_Q$ is split into three parts to generate the control, readout and stabilization drives. The first one is directly fed into the IQ mixer to generate the single-photon drive, which is connected to the weakly coupled port (orange) of the device. The second one is fed into a frequency doubler to make its frequency doubled to $\omega_S\approx2\omega_Q$ and generate the two-photon stabilization drive with the IQ mixer. Finally, the last one is mixed with the readout LO to generate the signal at the cat-qubit readout frequency $\omega_{CQR}\approx\omega_R-\omega_S/2$. The two-photon stabilization drive and the cat-qubit readout drive are both connected to the strongly coupled port (purple) of the device through a directional coupler.

The parameters of the device and the calibration/estimation methods are described in Tab. \ref{params}.

\begin{table}[!h]

\caption{\textbf{Device Parameters and Calibration Methods}}
\centering
\begin{tabular}{ |c| c |c|}
\hline 
 \textbf{Parameter} & \textbf{Value}  & \textbf{Methods}   \\
\hline    
SNAIL oscillator frequency $\omega_Q$& $2\pi\times5.9$ GHz & two-tone spectroscopy    \\ 
\hline    
SNAIL oscillator Kerr $K$& $2\pi\times1.2$ MHz & two-tone spectroscopy    \\ 
\hline 
SNAIL large junction $E_J$& $2\pi\times263.2$ GHz & room-temp. resistance msmt.\cite{ambegaokar1963tunneling}    \\
\hline 
SNAIL small-to-large junction ratio $\beta$ & 0.1 & room-temp. resistance msmt.\cite{ambegaokar1963tunneling}    \\
\hline
SNAIL oscillator charge energy $E_c$ & $2\pi\times$ 118 MHz & finite-element-method simulation    \\
\hline
SNAIL oscillator relaxation time $T_1$ & 40 $\mathrm{\mu}s$ & time-domain coherence msmt.  \\
\hline
SNAIL oscillator dephasing time $T_2$ & 3.2 $\mathrm{\mu}s$ & time-domain Ramsey msmt.\cite{steck2007quantum}  \\
\hline
SNAIL oscillator thermal population $n_\text{th}$ & 5\% & $T_z$ fitting with theoretical model  \\
\hline
\hline
Readout resonator frequency $\Omega_R$ & $2\pi\times$7.1 GHz & microwave reflectrometry    \\
\hline
Readout resonator relaxation rate $\kappa_R$ & $2\pi\times$0.4 MHz & microwave reflectrometry    \\
\hline
Readout resonator dispersive shift $\chi$ & $2\pi\times$40 kHz & time-domain spectroscopy    \\
\hline
\end{tabular}
\label{params}
\end{table}

\section{KCQ Hamiltonian Parameter Calibration}

The KCQ Hamiltonian consists of three parameters: the detuning $\Delta = \omega_Q - \omega_S/2$, two-photon stabilization drive strength $\epsilon_2$ and Kerr coefficient $K$. The parameters are extracted from a series of measurements. Without losing generality, we can set $\epsilon_2$ to be real, and the two-photon stabilization drive frequency $\omega_S$ is known from the hardware.

\begin{figure}[h!]
    \centering
    \includegraphics[width=0.5\textwidth]{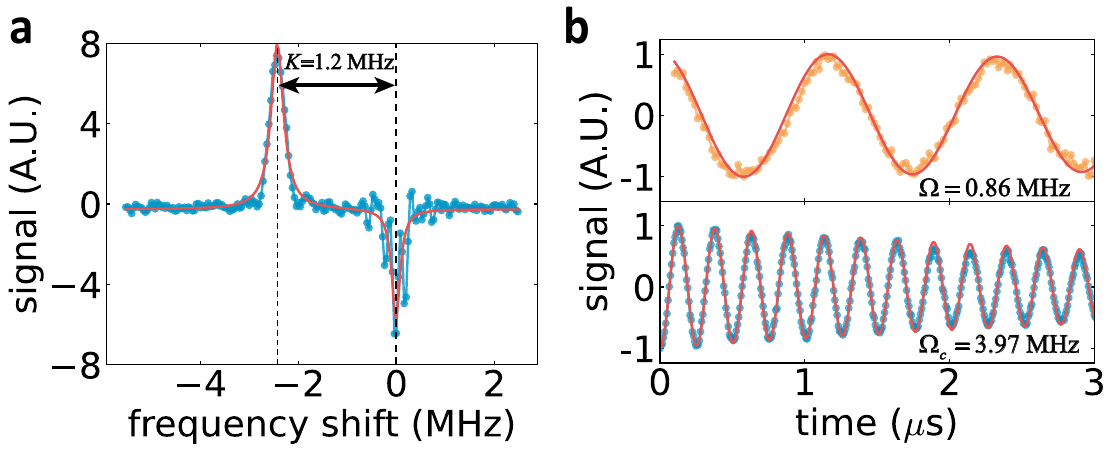}
    \caption{KCQ Hamiltonian calibration. \textbf{a} Calibration of the Kerr nonlinearity. \textbf{b} Calibration of the two-photon stabilization drive. }
    \label{fig:5}
\end{figure}

We first extract the KCQ nonlinear oscillator parameters without the two-photon stabilization drive, i.e. $\epsilon_2=0$. The KCQ nonlinear oscillator frequency $\omega_Q=\omega_{01}$ is the transition frequency between $|0\rangle$ and $|1\rangle$, determined using standard two-tone spectroscopy\cite{krantz2019quantum}. The Kerr nonlinearity is then extracted by preparing the nonlinear oscillator in the first excited states $|1\rangle$ followed by a similar two-tone spectroscopy. where we observed two peaks corresponding to the $|1\rangle \rightarrow|0\rangle$ transition with frequency $\omega_{01}$ and $|1\rangle \rightarrow|2\rangle$ transition with frequency $\omega_{12} = \omega_{01} - 2K$, which allows us to extract the Kerr nonlinearity $K=1.2~\text{MHz}$, as shown in Fig. \ref{fig:5}a.

Subsequently, we apply the two-photon stabilization drive to create cat states in the SNAIL nonlinear oscillator, and extract the value of $\epsilon_2$. With the two-photon stabilization drive, the KCQ Hamiltonian supports cat states $|C_\alpha^\pm\rangle \propto |+\alpha\rangle\pm|-\alpha\rangle$ with mean photon number $|\alpha|^2 = (\epsilon_2+\Delta/2)/K$. Because $K$ and $\Delta$ are already calibrated, we only need to calibrate $|\alpha|^2$. As mentioned in reference\cite{grimm2020stabilization, hajr2024high}, a single photon drive $H_d = \frac{\Omega}{2} \hat{a}^\dagger + \frac{\Omega^*}{2}\hat{a}$ induces a transition between the two cat states with the rate $\Omega_c = 2\text{Re}[\Omega]\alpha$ with $\epsilon_2>0$. Conversely, without the two-photon stabilization drive ($\epsilon_2=0$), the same drive can induce the transition between $|0\rangle$ and $|1\rangle$ states of the nonlinear oscillator with the rate $\Omega$. Therefore, by tuning the phase of the single photon drive to make $\epsilon_1$ real and measuring the two transition rates, we can extract, $\alpha = \Omega_c/(2\text{Re}[\Omega])$, as shown in Fig. \ref{fig:5}b.

\section{Detuned-KCQ Continuous $Z(\theta)$ Gates Calibration}

\begin{figure}[h!]
    \centering
    \includegraphics[width=0.66\textwidth]{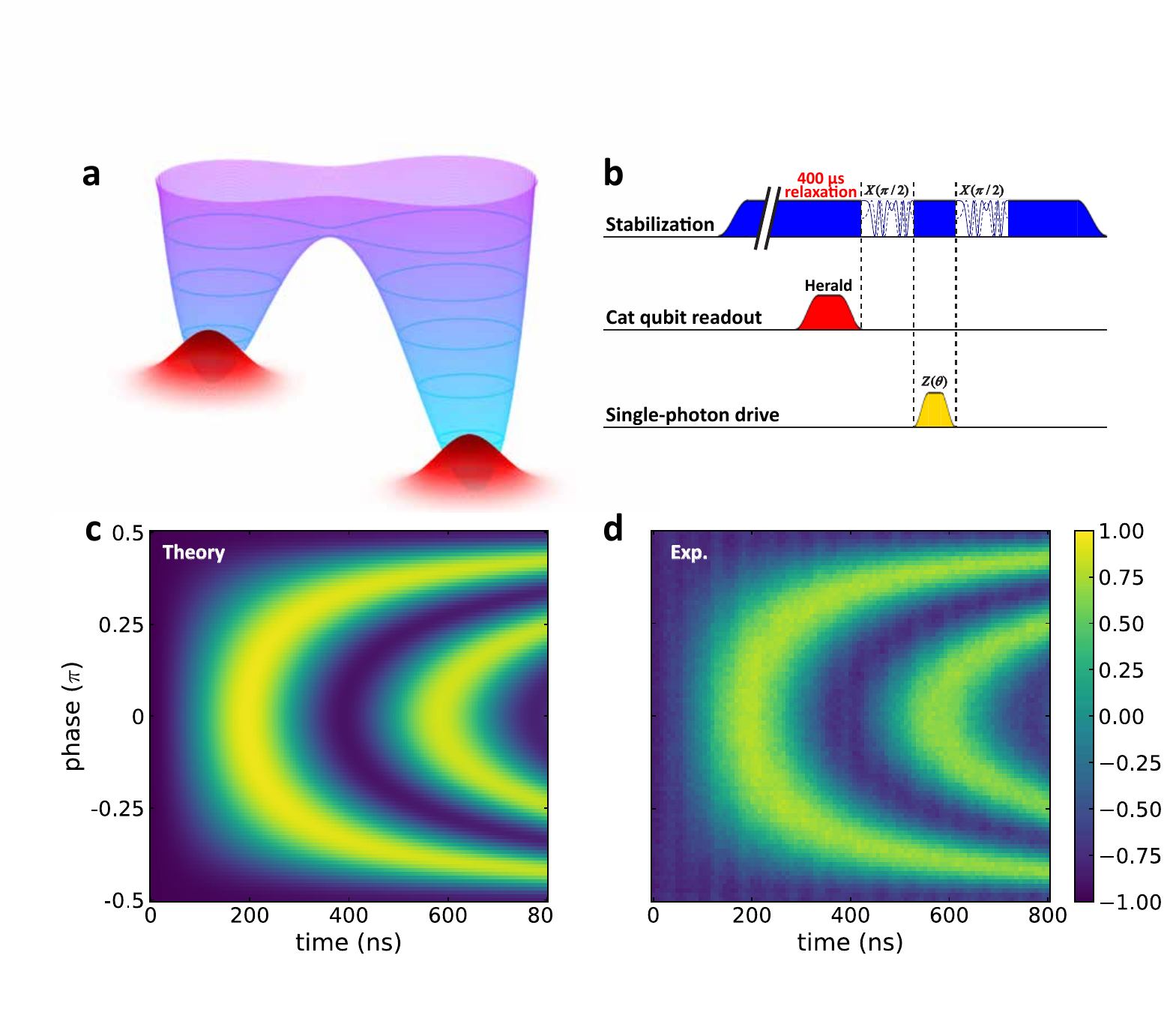}
    \caption{Calibration of $Z(\theta)$ gates. \textbf{a} The shifted energy profile when applying single-photon drive. \textbf{b} The sequence for calibration the $Z(\theta)$ gates. \textbf{c} Simulation of the measured Chevron-like plot. \textbf{d} The measured Chevron-like plot with the sequence above.}
    \label{fig:Z}
\end{figure}

The continuous $Z(\theta)$ gates is engineered by a single-photon drive at half of the stabilization drive frequency, which shifts the energy of the two pseudo-potential well, as shown in Fig. \ref{fig:Z}a. We calibrate $Z(\theta)$ gates following the pulse sequence shown in Fig. \ref{fig:Z}b. The pre-calibrated $X(\pi/2)$ gates are used to prepare and measure the state along Y axis. The phase and duration of the single-photon drive are swept to obtain the Chevron-like plot shown in Fig. \ref{fig:Z}d, which is in good agreement with the numerical simulation shown in Fig. \ref{fig:Z}c. The $Z(\theta)$ gate speed can be increased by increasing either the cat qubit size or the single-photon drive strength. However, the large single-photon drive strength will lead to non-adiabatic error discussed in the Main Text.

\section{Device Fabrication}

\begin{figure}[h!]
    \centering
    \includegraphics[width=0.93\textwidth]{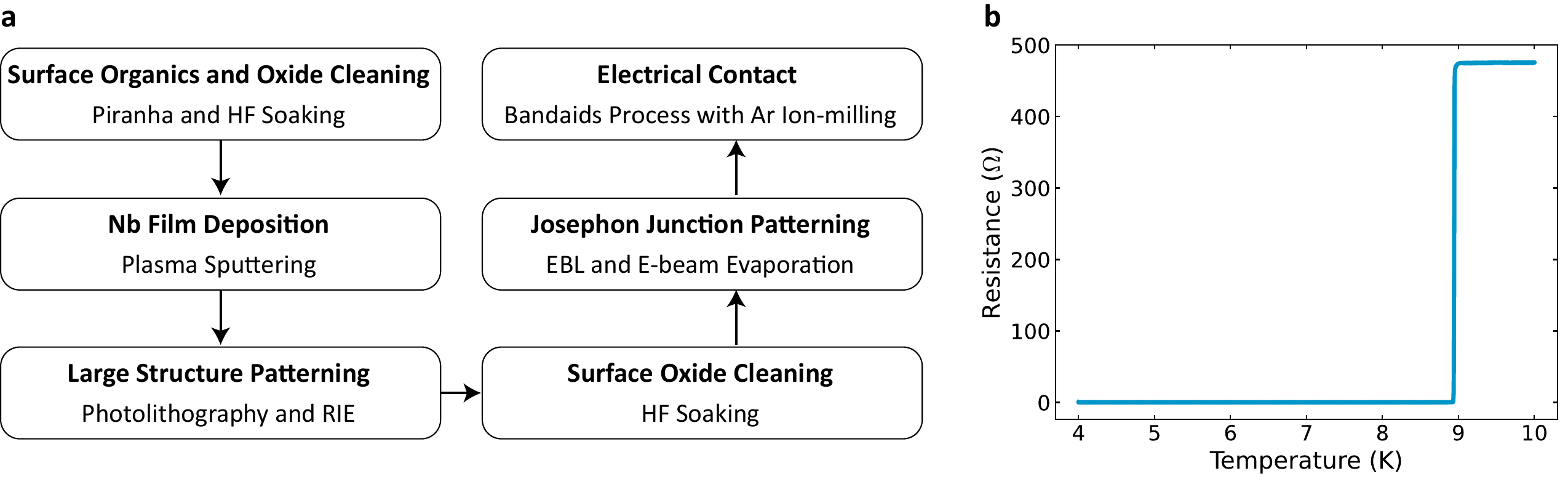}
    \caption{Fabrication and film quality. \textbf{a} Flow chart of the device fabrication process. \textbf{b} Measurement of the superconducting transition temperature of the Nb film. }
    \label{fig:Tc}
\end{figure}

The device in this work is based on superconducting circuits patterned on an intrinsic silicon (Si) wafer with a resistivity $\rho \geq 10 \mathrm{k \Omega \cdot cm}$. The fabrication process is briefly depicted in the flow chart shown in Fig. \ref{fig:Tc}a. Before the metal deposition, the wafer is cleaned by Piranha solutions ($\mathrm{H_2SO_4+H_2O_2}$) and hydrofluoric acid (HF) to remove the organic contamination and silicon oxide on the surface. The wafer is then transferred to a vacuum box to a Kert J. Lesker sputtering tool for niobium (Nb) deposition. By adjusting the sputtering temperature and pressure, a uniform niobium layer with compressive stress and high superconducting transition temperature\cite{kuroda1988niobium} is sputtered onto the wafer. The superconducting transition temperature of the film is measured to be 8.94 K, as shown in Fig. \ref{fig:Tc}b. The large structures, including the microwave buses, filter, quarter-wavelength resonators and capacitor pads, are patterned on the Nb film using photolithography and reactive ion etching (RIE). After a second cleaning process with HF to further remove oxides on Si and Nb surfaces, the Josephson junctions are defined with e-beam lithography (EBL) and bridge-free two-angle shadow evaporation of aluminum (Al) films\cite{lecocq2011junction}. Electrical contact between the Al Josephson junctions and the Nb capacitor pads is formed by the bandaid process with argon ion-milling\cite{qing2024broadband}. Once diced, the device is packaged into a oxygen-free copper box for testing in the dilution refrigerator.

\section{GST Circuit Design}

GST is a self-consistent calibration free protocol for characterizing the noisy implementation of a informationally complete set of quantum gates\cite{nielsen2021gate, brieger2023compressive, cao2024efficient}. GST can estimate the Pauli transfer matrix (PTM) of quantum gates in a SPAM-free manner, and is capable of fitting models with physical constraints such as completely positive, trace preserving (CPTP) constraints\cite{nielsen2021gate, brieger2023compressive, cao2024efficient}. This technique has been applied to various platforms of quantum computing, including ion traps\cite{blume2013robust}, spin qubits\cite{stemp2023tomography, bartling2024universal} and superconducting circuits\cite{hashim2023benchmarking}. GST usually executes a set of circuits that consist of a state preparation and readout sequence (\emph{fiducials}) and repeated sequences of gate operations (\emph{germs}) that amplify specific types of error \cite{nielsen2019python, nielsen2021gate}. 

Because the local gates of KCQ are the discrete $X(\pi/2)$ gate and continuous $Z(\theta)$ gates, we construct the informationally complete set of quantum gates as $\{X(\pi/2), Z(\pi/2), Y(\pi/2)\}$, where the $Y(\pi/2)$ gate is complied using local gates as $X(\pi)Z(\pi/2)X(\pi/2)Z(\pi/2)$. Each GST circuit has the form $F_i G^l F_j$, where $F_i$ and $F_j$ are \emph{fiducial sequences}, $G$ is a \emph{germ} sequence, and $l$ is the maximum integer such that the length of $G^l$ is at most some maximum length $L$. Using the pyGSTi python package\cite{nielsen2019pythonS}, we generate the GST circuits which have a germ set with $10$ germs of maximum length $5$. The fiducials and germs are shown in Tab. \ref{GST_circuit}. 

\begin{table}[!ht]

\caption{\textbf{Gate Set Tomography Germs and Fiducials}}
\centering
\begin{tabular}{ |c  | l |}
\hline
\centering
Germs &  $Z(\pi/2), X(\pi/2)$, $Y(\pi/2)$, $Y(\pi/2)X(\pi/2)$, $X(\pi/2)Z(\pi/2)$,\\ 
& $Y(\pi/2)Z(\pi/2$), $Y(\pi/2)X(\pi/2)X(\pi/2)Z(\pi/2)$, \\
& $Y(\pi/2)Y(\pi/2)Y(\pi/2)Z(\pi/2)$, $X(\pi/2)Z(\pi/2)Y(\pi/2)Z(\pi/2)Z(\pi/2)$, \\
& $X(\pi/2)X(\pi/2)X(\pi/2)Z(\pi/2)$ \\
\hline    
Preparation Fiducials & $X(\pi/2)$, $Y(\pi/2)$, $X(\pi/2)X(\pi/2)$, no gates    \\ 
\hline
Measurement Fiducials & $X(\pi/2)$, $Y(\pi/2)$, $X(\pi/2)X(\pi/2)$, no gates \\
\hline
\end{tabular}
\label{GST_circuit}
\end{table}

We perform GST with 800 circuits of maximum lengths $L \in \{0,1,2,4,8,16,32,64,128\}$ and each circuit is implemented with 1024 shots to obtan the statistics of the measurement results. The process matrices and error generators of $X(\pi/2)$ and $Z(\pi/2)$ gates are extracted by fitting the measurement results to a model with CPTP constraints. 

However, the uncertainty of the extracted Pauli $X$ and Pauli $Y$ errors from GST is very large due to the shallow depth of the GST circuits (128 at maximum), which hinders the exact characterization of the noise-bias. Therefore, we employ the DRB protocol with a maximum circuit depth of 2000 to explore the noise structure of KCQ and demonstrate that the performance of detuned-KCQ crosses the fault-tolerant threshold of the XZZX surface code\cite{darmawan2021practical, claes2023estimating}.

\section{$\mathbb{D}_8$ Dihedral Group Randomized Benchmarking Design}

The bit-flip error and phase-flip error, as well as the noise bias are extracted by the DRB protocol introduced in references\cite{carignan2015characterizing, claes2023estimating} with a scaling factor to compensate the effect of noise-free $X(\pi)$ gate. We will describe the protocol and illustrate the extraction of the scaling factor by numerical simulations.

\begin{figure}[h!]
    \centering
    \includegraphics[width=0.5\textwidth]{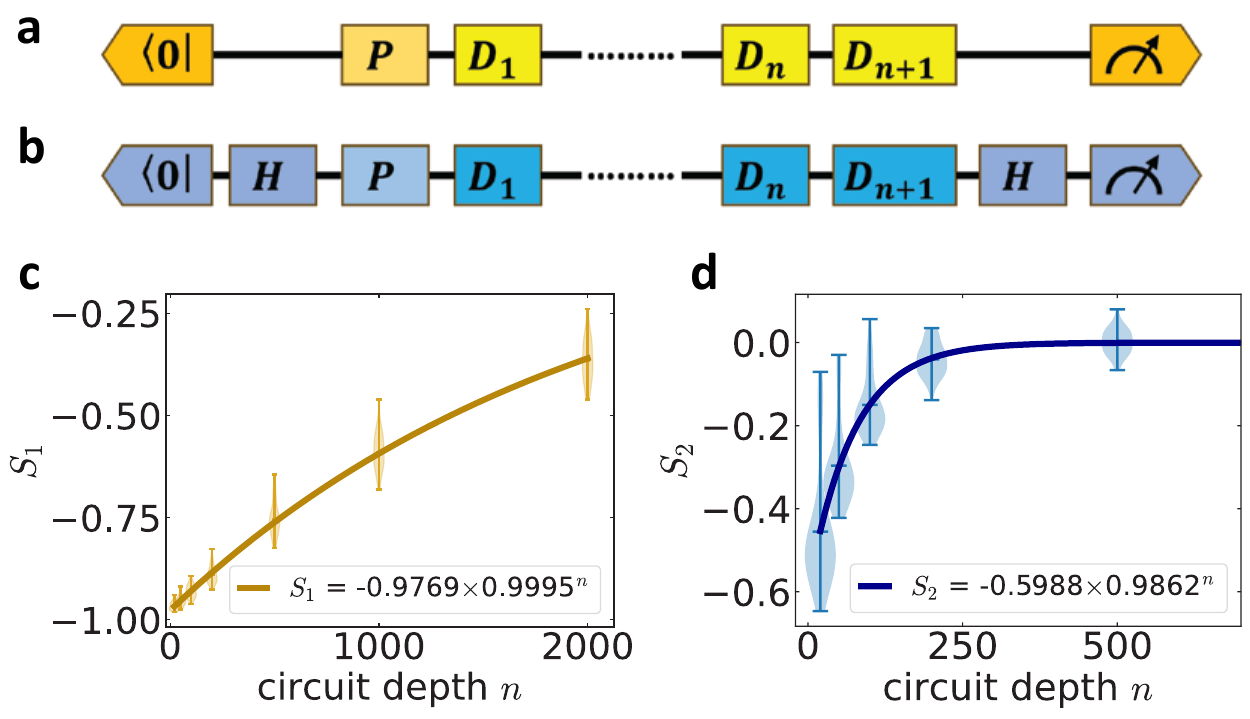}
    \caption{DRB survival probability fitting. Circuits and fitting curve examples of the exponential decay in the $\mathbb{D}_8$ dihedral randomized benchmarking protocol when \textbf{a, c} preparing and measuring the states in $Z$ axis, and \textbf{b, d} preparing and measuring the states in $X$ axis. The error bars indicate the shot noise from 1024 repeated measurements.}
    \label{fig:8}
\end{figure}

The protocol consists of two modified RB sequences labeled by $b = 1$ and $2$, shown in Fig. \ref{fig:8}a and Fig. \ref{fig:8}b, which prepare and measure the qubit along $Z$ and $X$ axis, respectively. The random gates in RB sequences consist of a gate $P$ sampled from the single qubit Pauli group $\mathbb{P}$, followed by $n$ gates $D_k$ sampled from the $\mathbb{D}_8$ dihedral group, and a final gate $D_{n+1} = \left(D_n...D_2D_1\right)^{-1}$ to reverse the states. The bit-flip error and phase-flip error are extracted by fitting the character-weighted survival probability $S_b(n)$ (shown in Eq. \ref{Sbn}) to an exponential function.

\begin{equation}
    S_b(n):=\underset{\substack{P \in \mathbb{P} \\ U_1 \cdots U_n \in \mathbb{D}_8}}{\mathsf{E}}\left[\chi_b^*\left(P\right) \Gamma_{\left\{U_i\right\}}\right]
    \label{Sbn}
\end{equation}
Here, $\Gamma_{\{U_i\}}$ represents the expectation value of the measurement after applying a sample of RB sequences, while $\mathsf{E}[\cdot]$ denotes the average over all samples of gates. A few typical fitting plots are shown in Fig. \ref{fig:8}c and Fig. \ref{fig:8}d.

\begin{figure}[h!]
    \centering
    \includegraphics[width=0.5\textwidth]{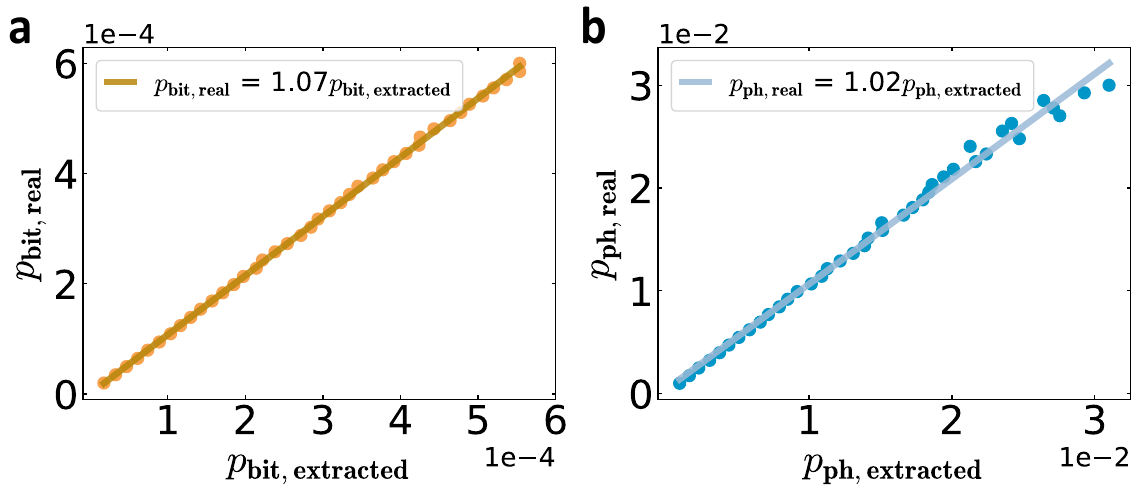}
    \caption{Scaling factor \textbf{a} between the extracted and real bit-flip errors, and \textbf{b} between the extracted and real phase-flip errors}
    \label{fig:9}
\end{figure}

The scaling factor between the errors of $Z(\theta)$ rotations and the extracted errors from DRB protocol is derived by numerical simulation with qiskit\cite{wille2019ibm} package. We introduce various phase-flip errors $p_\text{pf, \text{real}}$ ranging from $0$ to $3\%$ and bit-flip errors $p_\text{bf, real}=0.02p_\text{pf, real}$ to $Z(\theta)$ rotations, while keeping the $X(\pi)$ gate noise-free, and then simulate the DRB experiments to extract the bit-flip $p_\text{pf, extracted}$ and phase-flip errors $p_\text{pf, extracted}$. As shown in Fig. \ref{fig:9}, the extracted errors are proportional to the errors on the $Z(\theta)$ rotations. By fitting the simulation results with a linear function, we extracted the scaling factor between them, where $p_\text{pf, real} = 1.02p_\text{pf, extracted}$ and $p_\text{bf, real} = 1.07p_\text{bf, extracted}$.

\section{Description and Analysis of the Error Channels}
We model the error channels of KCQ as a combination of coherent Pauli rotation errors and incoherent stochastic Pauli errors described by a Pauli transfer matrix $\mathcal{E} = e^\mathbf{L}$ with the generator $\mathbf{L} = h_x \mathbf{H_x}+h_y \mathbf{H_y}+h_z \mathbf{H_z}+
p_x \mathbf{P_x}+p_y \mathbf{P_y}+p_z \mathbf{P_z}$. The coherent components in the error generator are defined in Eq. \ref{coherent}, and the incoherent components are defined in Eq. \ref{incoherent}.\cite{blume2022taxonomy, carignan2024estimating}

\begin{equation}
    H_x = 
    \begin{pmatrix}
0 & 0 & 0 & 0 \\
0 & 0 & 0 & 0 \\
0  & 0 & 0 & -1  \\
0 & 0 & 1 & 0 
\end{pmatrix},
    H_y = 
    \begin{pmatrix}
0 & 0 & 0 & 0 \\
0 & 0 & 0 & 1 \\
0  & 0 & 0 & 0  \\
0 & -1 & 0 & 0 
\end{pmatrix},
    H_z = 
    \begin{pmatrix}
0 & 0 & 0 & 0 \\
0  & 0 & -1 & 0  \\
0 & 1 & 0 & 0 \\
0 & 0 & 0 & 0 
\end{pmatrix}
\label{coherent}
\end{equation}

\begin{equation}
    P_x = 
    \begin{pmatrix}
0 & 0 & 0 & 0 \\
0 & 0 & 0 & 0 \\
0  & 0 & -2 & 0 \\
0 & 0 & 0 & -2 
\end{pmatrix},
    P_y = 
    \begin{pmatrix}
0 & 0 & 0 & 0 \\
0 & -2 & 0 & 0 \\
0  & 0 & 0 & 0  \\
0 & 0 & 0 & -2 
\end{pmatrix},
    P_z = 
    \begin{pmatrix}
0 & 0 & 0 & 0 \\
0  & -2 & 0 & 0  \\
0 & 0 & -2 & 0 \\
0 & 0 & 0 & 0 
\end{pmatrix}
\label{incoherent}
\end{equation}

In the limit of small errors, we can expand the Pauli transfer matrix as $\mathcal{E} = 1 + \mathbf{L} + 0.5\mathbf{L}^2 + o(\mathbf{L}^3)$ in Eq. \ref{error_expand} up to $O(p_m), O(h^2_n)$ with $m,n \in \{x, y, z\}$ and $p = p_x+p_y+p_z$, as shown below

\begin{equation}
    \begin{pmatrix}
        0 & 0 & 0 & 0 \\
        0 & 1-2(p_y+p_z)-\frac{1}{2}(h_y^2+h_z^2) & -h_z(1+p)+h_xh_y & h_y(1+p) + 2h_xh_z \\
        0 & h_z(1+p) + h_xh_y & 1-2(p_x+p_z)-\frac{1}{2}(h_x^2+h_z^2) & -h_x(1+p) + h_yh_z \\
        0 & -h_y(1+p)+h_xh_z & h_x(1+p)+h_yh_z & 1-2(p_x+p_y)-\frac{1}{2}(h_x^2+h_y^2) \\
    \end{pmatrix}
    \label{error_expand}
\end{equation}

The coherent error manifests as the off-diagonal elements of the Pauli transfer matrix, which can be effectively removed by the Pauli twirling process. The Pauli twirling process randomly insert Pauli gates $P\in\mathbb{P}$ while compiling circuits, and all the Pauli gates can be executed with error-free virtual $X$ gate, high-fidelity bias-preserving $Z$ gate or their combination. Therefore, Pauli twirled error channels introduces minimal extra noise with preserved the noise-bias property, and are reported to be statistically the same as the full error channels in $\mathtt{XZZX}$ surface code\cite{darmawan2021practical}.

The Pauli transfer matrix of the error process after Pauli twirling process reduced to that of a Pauli stochastic error channel shown in Eq. \ref{error_reduced}, 

\begin{equation}
    \mathcal{E} = 
    \begin{pmatrix}
        0 & 0 & 0 & 0 \\
        0 & 1-2(p'_y+p'_z) & 0 & 0 \\
        0 & 0 & 1-2(p'_x+p'_z) & 0 \\
        0 & 0 & 0 & 1-2(p'_x+p'_y) \\
    \end{pmatrix}
    \label{error_reduced}
\end{equation}
where the associated stochastic Pauli errors are $p'_m = p_m + 0.25h_m^2, m\in\{x,y,z\}$. Even with the highest coherent error rate, $h_x = 0.0038\pm0.002$, of the bias-preserving $Z(\pi/2)$ gate estimated from GST, the contribution of stochastic Pauli $X$ error from this coherent error is $3.61\times10^{-6}\pm{3\times10^{-6}}$, which is two orders or magnitude smaller than the stochastic Pauli $X$ error. Therefore, the effects of coherent errors are negligible. 



\end{document}